\documentclass[modern]{aastex62}

\usepackage{amsmath}

\graphicspath{{./}{figures/}}

\received{XXXXX}
\revised{XXXXX}
\accepted{\today}
\submitjournal{ApJ}

\shorttitle{Stellar Metallicities for the CKS Sample}
\shortauthors{Ghezzi et al.}

\begin{document}

\title{A SPECTROSCOPIC ANALYSIS OF THE CALIFORNIA-$KEPLER$ SURVEY SAMPLE: II. CORRELATIONS OF STELLAR METALLICITIES WITH PLANETARY ARCHITECTURES}

\correspondingauthor{Luan Ghezzi}
\email{luanghezzi@astro.ufrj.br}

\author[0000-0002-9089-0136]{Luan Ghezzi}
\affiliation{Universidade Federal do Rio de Janeiro, Observat\'orio do Valongo, Ladeira do Pedro Ant\^onio, 43, Rio de Janeiro, RJ 20080-090, Brazil}

\author[0000-0002-0786-7307]{Cintia F. Martinez}
\affiliation{Observat\'orio Nacional, Rua General Jos\'e Cristino, 77, Rio de Janeiro, RJ 20921-400, Brazil}

\author[0000-0002-4235-6369]{Robert F. Wilson}
\affiliation{Department of Astronomy, University of Virginia, Charlottesville, VA 22904-4325, USA}

\author[0000-0001-6476-0576]{Katia Cunha}
\affiliation{Observat\'orio Nacional, Rua General Jos\'e Cristino, 77, Rio de Janeiro, RJ 20921-400, Brazil}
\affiliation{Steward Observatory, University of Arizona, 933 North Cherry Avenue, Tucson, AZ 85721-0065, USA}

\author[0000-0002-0134-2024]{Verne V. Smith}
\affiliation{NSF's NOIRLab, 950 North Cherry Avenue, Tucson, AZ 85719, USA}

\author[0000-0003-2025-3147]{Steven R. Majewski}
\affiliation{Department of Astronomy, University of Virginia, Charlottesville, VA 22904-4325, USA}



\begin{abstract}
We present independent and self-consistent metallicities for a sample of 807 planet-hosting stars from the California-Kepler Survey from an LTE spectroscopic analysis using a selected sample of Fe I and Fe II lines. 
Correlations between host-star metallicities, planet radii, and planetary architecture (orbital periods  -- warm or hot -- and multiplicity -- single or multiple), were investigated using non-parametric statistical tests.
In addition to confirming previous results from the literature, e.g., that overall host star metallicity distributions differ between hot and warm planetary systems of all types, we report on a new finding that when comparing the median metallicities of hot versus warm systems, the difference for multiple Super-Earths is considerably larger when compared to that difference in single Super-Earths. 
The metallicity CDFs of hot single Super-Earths versus warm single Super-Earths indicate different parent stellar populations, while for Sub-Neptunes this is not the case.
The transition radius between Sub-Neptunes and Sub-Saturns was examined by comparing the APOGEE metallicity distribution for the Milky Way thin disk in the solar neighborhood with metallicity distributions of host stars segregated based upon the largest known planet in their system. 
These comparisons reveal increasingly different metallicity distributions as the radius of the largest planet in the systems increases, with the parent stellar metallicities becoming significantly different for R$_{p}>$ 2.7 R$_{\oplus}$.  The behavior of the p-values as a function of planet radius undergoes a large slope change at R$_{p}$ = 4.4 $\pm$ 0.5 R$_{\oplus}$, indicating the radius boundary between small and large planets.
\end{abstract}


\section{Introduction} 
\label{sec:intro}

Shortly after the discovery of the first exoplanet orbiting a solar-type star \citep{mayor1995}, \cite{gonzalez1997} presented intriguing evidence that stellar metallicities may play a central role in the process of planetary formation. Subsequent studies with ever increasing samples confirmed that FGK main-sequence and subgiant stars hosting giant planets have higher metallicities, on average, when compared to their counterparts not known to harbor similar planets \citep[e.g.,][]{santos2004, fischer2005, ghezzi2010a, sousa2011}. It seems that the formation of giant planets is favored around more metal-rich stars \citep[e.g.,][]{fischer2005,johnson2010a,ghezzi2018}, a result that agrees with predictions from the core accretion model \citep{ida2004}.

Although more than 4,400 planets orbiting more than 3,200 stars are known as of June 2021 (NASA Exoplanet Archive\footnote{https://exoplanetarchive.ipac.caltech.edu/} and The Extrasolar Planets Encyclopaedia\footnote{http://exoplanet.eu/}), answers to many questions regarding the connection between stellar metallicity (and other chemical abundances) and planetary formation remain elusive. For instance, the very origin of the so-called planet-metallicity correlation is not yet explained and it is not clear if this relation also holds for giant stars or smaller planets \citep[e.g.,][]{mulders2018,adibekyan2019}. Of particular interest is whether the formation of terrestrial planets depends on stellar metallicity, which has significant implications for the distribution of habitable planets (and maybe life) in the Milky Way and the Universe.

As the increased precision of the Doppler surveys allowed the first discoveries of Neptunian-mass and smaller planets (i.e., $M_{p}$ sin $i$ $\lesssim$ 30 M$_{\oplus}$), \cite{sousa2008} showed that their host stars had lower average metallicities than those orbited by Jovian planets. Using a somewhat larger (but still small) sample, \cite{ghezzi2010a} confirmed this was valid for stars hosting only Neptunian-mass or smaller planets. By the time the latter study was published, the {\it Kepler} mission \citep{borucki2010} was already detecting its first planets. During its operation, {\it Kepler} discovered more than 2,300 planets, with the majority having $R_{p}$ $<$ 4.0 R$_{\oplus}$, i.e., radii smaller than that of Neptune (NASA Exoplanet Archive). The {\it Kepler} sample thus provides a unique opportunity to investigate if and how stellar metallicity influences the formation and architectures of planetary systems. 

Initial results revealed that smaller planets, unlike gas giants, are found around stars with a wide range of metallicities, with an average close to solar \citep{buchhave2012,everett2013}. Moreover, \cite{buchhave2014} suggested that metallicity might play a fundamental role in defining the structures of the planets in a given system, segregating them into three different regimes: terrestrial ($R_{p}$ $<$ 1.7 R$_{\oplus}$), gas dwarfs (1.7 R$_{\oplus}$ $<$ $R_{p}$ $<$ 3.9 R$_{\oplus}$) and gas giants ($R_{p}$ $>$ 3.9 R$_{\oplus}$). \cite{schlaufman2015} argued against the existence of the boundary at 1.7 R$_{\oplus}$, and \cite{wang2015} proposed that the planet-metallicity correlation is universal, extending from gas giants down to terrestrial planets. However, a subsequent analysis by \cite{buchhave2015} provided further support to the idea that the planet-metallicity correlation is not valid for small planets. 

The role of stellar metallicity was further explored by \cite{mulders2016}, who used spectroscopic metallicities for more than 20,000 {\it Kepler} stars from the LAMOST survey (as part of the LAMOST-{\it Kepler} project ---  \citealt{decat2015,frasca2016}) and found that planets with periods shorter than 10 days preferentially orbit metal-rich stars. \cite{wilson2018} found a similar correlation between [Fe/H] and orbital period using SDSS-IV APOGEE \citep{majewski2017} metallicities and determined a critical period $P$ $\sim$ 8 days below which small planets ($R_{p}$ $\leq$ 4.0 R$_{\oplus}$) orbit more metal-rich stars. These results received further support from \cite{petigura2018}, who analyzed metallicities from the California-{\it Kepler} Survey (CKS; \citealt{petigura2017a,johnson2017}) to conclude that planet occurrence increases with [Fe/H] for hot planets of all sizes, although with different strengths for each planet subclass. For warm Super-Earths (1.0 $\leq$ $R_{p}$ $\leq$ 1.7 R$_{\oplus}$), on the other hand, planet occurrence is not correlated with metallicity. \cite{dong2018} used LAMOST stellar parameters to focus on a population of short-period, Neptune-sized planets (1 d $<$ $P$ $<$ 10 d and 2 R$_{\oplus}$ $\lesssim$ $R_{p}$ $\lesssim$ 6 R$_{\oplus}$), which led to a similar conclusion as previous studies: Neptunes with short periods orbit more metal-rich stars.

\cite{martinez2019} (hereafter Paper I) determined homogeneous spectroscopic parameters (stellar effective temperatures, surface gravities, and radii) from an analysis of a sample of 1232 stars from the CKS. Combining their effective temperatures with Gaia distances, the authors presented stellar radii with a median uncertainty of 2.8\% and computed radii for 1633 planets achieving an internal precision of 3.7\%, which revealed a slope of the radius gap as a function of orbital period. The present work continues the analysis performed in Paper I and presents the stellar metallicities (as Fe abundances) based upon the same classical spectroscopic methodology as in Paper I, which is completely independent from the methodology presented in CKS.

Using available results from CKS, \cite{munozromero2018} investigated if stellar metallicity could also influence planet multiplicity, but found no significant differences between [Fe/H] distributions for different categories of systems with singles and multiple transiting planets. Also using CKS parameters, \cite{weiss2018} found no correlation with metallicity for the regime of small transiting planets ($R_{p}$ $\leq$ 4.0 R$_{\oplus}$), and \cite{zhu2018} also found similar results for planets of all sizes based on LAMOST parameters. These results would suggest a common origin for Kepler systems with single or multiple planets. In this study, we use our independently derived set of parameters to further investigate correlations between host star metallicities and planetary architectures in terms of planet size, orbital period and multiplicity, as such comparisons are important to further constrain models of planet formation \citep[e.g.,][]{owen2018,swain2019,loyd2020,gupta2020,mordasini2020,venturini2020}. In order to achieve this, our study focuses on trends between metallicities of the host stars and their planetary systems, not only on trends with individual planets.

The paper is organized as follows. In Section \ref{sec:obs}, we briefly describe the high-resolution spectroscopic CKS data analyzed and the determination of the metallicities. Our results are shown and compared to other determinations in the literature in Section \ref{sec:results}. We discuss our findings in Section \ref{sec:discussion} and present our conclusions in Section \ref{sec:conclusions}.

\section{Analysis} 
\label{sec:obs}

\subsection{Sample \& Observations}

The sample and data analyzed in this work are the same as in Paper I. In summary, our sample was selected from the California Kepler Survey \citep[CKS;][]{petigura2017a}, a large observational campaign targeting stars identified as {\it Kepler} Objects of Interest (KOIs). The CKS sample consists of 1305 stars hosting planets, but as in Paper I, 122 stars were removed from the sample due to different reasons, such as non-convergence of the automatic analysis method and unavailable stellar radii or with errors larger than 10\%. The remaining 1183 stars host 1781 planetary candidates according to {\it Kepler} Data Release 25 \citep[DR25;][]{thompson2018}. After excluding candidates classified as false positives according to the flag \textit{koi\_pdisposition}, we obtained the 1633 planets listed in Table 2 from Paper I and a corresponding sample of 1005 stars; this is the complete sample. Finally, the ``clean'' sample from Paper I was obtained by removing planets with $\sigma(R_{p}$)/$R_{p} >$ 8\%, impact parameter $b$ $>$ 0.7, and $P$ $>$ 100 d. This sample is composed of 961 transiting planets around 663 stellar hosts.

The CKS team obtained high-resolution spectra for the target stars with the HIRES (High Resolution Echelle Spectrometer; \citealt{vogt1994}) spectrograph on the Keck I 10 m telescope (Mauna Kea, Hawaii). We analyzed their reduced spectra, which are publicly available\footnote{https://california-planet-search.github.io/cks-website/}. The resolution of the HIRES spectra analyzed is $R$ = $\lambda$/$\Delta\lambda \sim$ 60,000 and the wavelength coverage is almost complete in the interval 3640 \AA\ -- 7990 \AA\ \citep{petigura2017a}. The $S/N$ ratios per resolution element of the spectra for targets in our complete sample range from $\sim$10 to $\sim$300, with a median value of $\sim$60.

\subsection{Determination of Metallicities}
\label{sec:feh}

The metallicities [Fe/H], along with the other atmospheric parameters (effective temperature $T_{\rm eff}$, surface gravity $\log g$ and microturbulent velocity $\xi$) used in this work were all derived at the same time and in a homogeneous and self-consistent way as discussed in Paper I. The automated iterative process and pipeline was originally developed by \cite{ghezzi2010a} and updated by \cite{ghezzi2018}. Readers are referred to these three papers for additional details on the spectroscopic analysis. 

We present metallicities for 807 stars, a subset of the sample from Paper I. Note that in this study we rejected stars having median errors in the automatically measured equivalent widths that were greater than $\sim$12\%, which was used as the main threshold for the quality cut.  In addition, we also rejected stars having derived microturbulent velocities, $\xi$, larger than 1.7 km s$^{-1}$; such high values of $\xi$ are unexpected and come mostly from the presence of a correlation between equivalent width errors and $\xi$.  Inspection of the rejected stars finds that they are composed mainly of two groups: the first is characterized by low-quality spectra having S/N less than $\sim$40, while the second is composed of stars having larger values of projected rotational velocities, with $v$ sin $i \gtrsim$ 7 km s$^{-1}$.  This quality cut in the data ensured that the derived metallicities have internal errors less than 0.06 dex, with a median error of 0.02 dex, but, our clean sample now contains 561 stars (804 planets), instead of the original 663 hosts from Paper I. We present in Table \ref{tab:ews} the equivalent width measurements and the resulting individual Fe abundances obtained for all Fe I and Fe II lines used to determine the metallicities for our sample, while the derived metallicities and their respective uncertainties are presented in Table \ref{tab:feh}.  


\begin{deluxetable*}{lccccccc}[t!]
\tablecaption{Equivalent widths and individual Fe abundances.
\label{tab:ews}}
\tablecolumns{8}
\tablewidth{0pt}
\tablehead{
\colhead{Star} & \colhead{$\lambda$} & \colhead{Ion\tablenotemark{a}} & \colhead{$\chi$} & \colhead{$\log gf$} & \colhead{$\log \Gamma_{w}$\tablenotemark{b}} & \colhead{EW\tablenotemark{c}} & \colhead{A(Fe)} \\
\colhead{} & \colhead{(\AA)} & \colhead{} & \colhead{(eV)} & \colhead{} & \colhead{$(sN_{H})^{-1}$} & \colhead{(m\AA)} & \colhead{(dex)}}
\startdata
 K00001  &  5023.185  &  26.0  &  4.283  & -1.524  & -7.135  &     29.5  &     7.36  \\
  K00001  &  5025.303  &  26.0  &  4.284  & -1.919  & -7.140  &     20.3  &     7.53  \\
 K00001  &  5044.211  &  26.0  &  2.851  & -2.206  & -7.280  &     78.3  &     7.63  \\
 K00001  &  5054.642  &  26.0  &  3.640  & -2.087  & -7.599  &     35.3  &     7.46  \\
 K00001  &  5058.496  &  26.0  &  3.642  & -2.809  & -7.599  &     10.9  &     7.49  \\
 K00001  &  5067.149  &  26.0  &  4.220  & -1.068  & -7.187  &  \nodata  &  \nodata  \\
 K00001  &  5109.651  &  26.0  &  4.301  & -0.853  & -7.150  &     75.0  &     7.53  \\
 K00001  &  5159.057  &  26.0  &  4.283  & -0.932  & -7.175  &     72.1  &     7.55  \\
 K00001  &  5196.059  &  26.0  &  4.256  & -0.732  & -7.510  &     69.6  &     7.39  \\
 K00001  &  5197.935  &  26.0  &  4.301  & -1.608  & -7.175  &  \nodata  &  \nodata  \\
\nodata & \nodata & \nodata & \nodata & \nodata & \nodata & \nodata & \nodata \\ 
\enddata
\tablenotetext{a}{Fe I and Fe II ions are represented by the notations 26.0 and 26.1, respectively, following the format used by MOOG \citep{sneden1973}.}
\tablenotetext{b}{$\log \Gamma_{w}$ is the logarithm of the van der Waals damping constant at 10,000 K.}
\tablenotetext{c}{Equivalent widths for some lines could not be measured by ARES \citep{sousa2015} or were removed during the iterative process (see \citealt{ghezzi2018} for more details).}
\tablecomments{This table is published in its entirety in the machine-readable format. A portion is shown here for guidance regarding its form and content.}
\end{deluxetable*}


\section{Results}
\label{sec:results}

\subsection{The Stellar Metallicity Distribution}
\label{sec:feh_dist}

\begin{figure}
\plotone{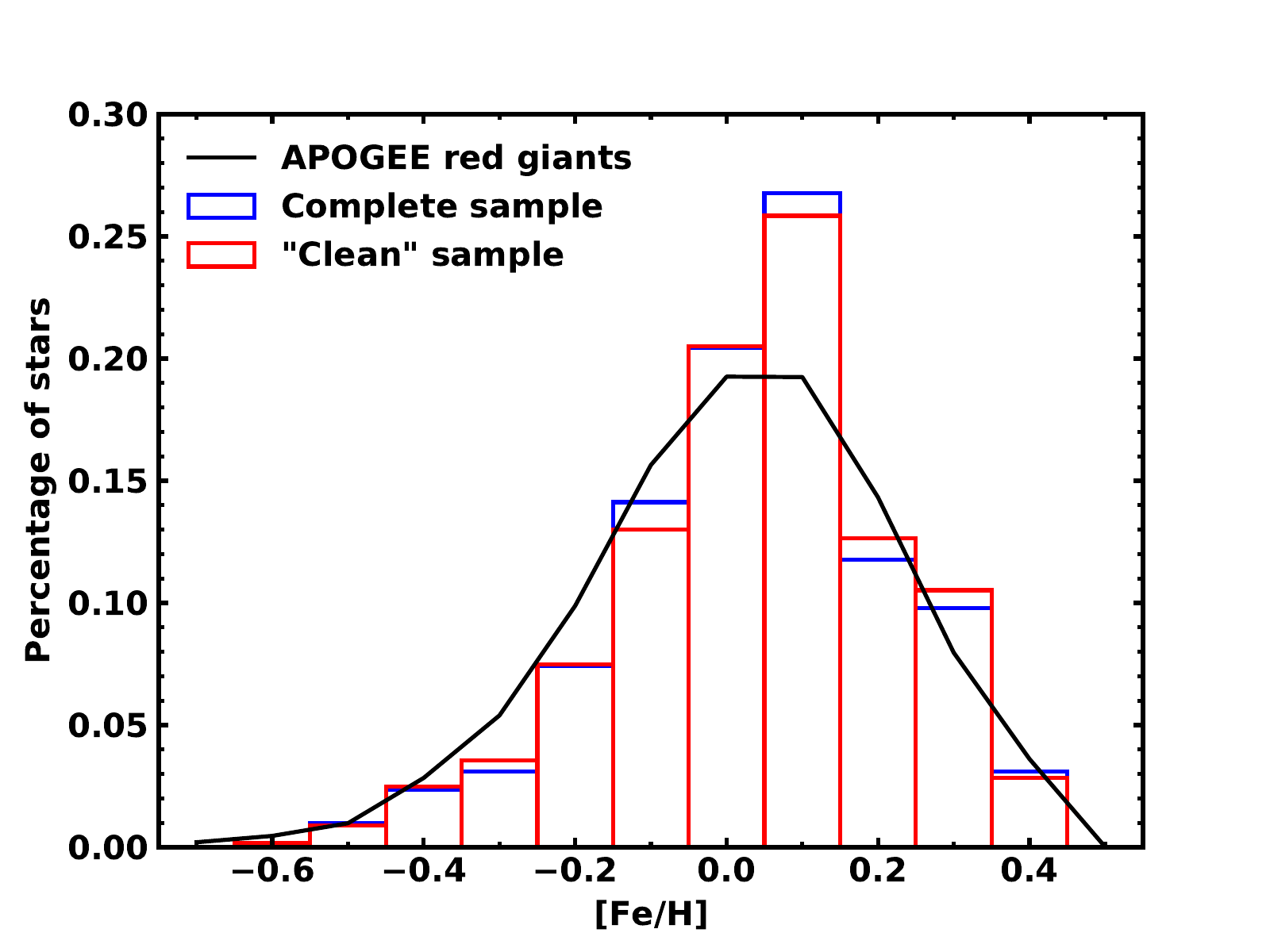}
\caption{Distributions of [Fe/H] for the complete (blue) and ``clean'' (red) samples of planet-hosting stars for which metallicities are from this study. The solid black line shows the metallicity distribution for disk stars in the solar neighborhood from the APOGEE Survey \citep{majewski2017}.}
\label{fig:feh_hist}
\end{figure}

The CKS sample is mainly composed of solar-type FGK main sequence stars (4700 K $\lesssim T_{\rm eff} \lesssim$ 6500 K and $\log g \gtrsim$ 4.0) but it also contains a smaller number of evolved stars ($\log g \lesssim$ 4.0; Paper I).
The metallicity distributions for both the complete and ``clean'' samples are shown in Figure \ref{fig:feh_hist} for comparison. It is clear that both distributions are very similar, which indicates that the cuts performed to obtain the ``clean'' sample have not introduced any significant biases. Both samples cover the range -0.60 $\lesssim$ [Fe/H] $\lesssim$ +0.44, with a peak slightly above solar metallicity ($\sim$0.10 dex), and 
are non-Gaussian distributions according to a Shapiro-Wilk test ($p_{SW} \sim 10^{-7} - 10^{-8}$). We note that we adopt the limit $p = 0.01$ in order to accept or reject the null hypotheses of the statistical tests performed throughout the paper. The median metallicity for both the complete and ``clean'' samples are the same: +0.06 $\pm$ 0.14 (median absolute deviation; MAD) dex. 


\begin{deluxetable*}{lrc}[t!]
\tablecaption{Stellar metallicities.
\label{tab:feh}}
\tablecolumns{3}
\tablewidth{0pt}
\tablehead{
\colhead{Star} & \colhead{[Fe/H]} & \colhead{$\sigma$([Fe/H])} \\
\colhead{} & \colhead{(dex)} & \colhead{(dex)}}
\startdata
 K00001  &  0.00  &  0.02  \\
 K00007  &  0.15  &  0.01  \\
 K00017  &  0.36  &  0.01  \\
 K00020  &  0.08  &  0.01  \\
 K00022  &  0.22  &  0.02  \\
 K00041  &  0.07  &  0.01  \\
 K00046  &  0.41  &  0.02  \\
 K00049  & -0.08  &  0.03  \\
 K00063  &  0.16  &  0.01  \\
 K00064  &  0.04  &  0.02  \\
\nodata & \nodata & \nodata \\ 
\enddata
\tablecomments{This table is published in its entirety in the machine-readable format. A portion is shown here for guidance regarding its form and content.}
\end{deluxetable*}



The metallicity distribution obtained for the CKS sample in this study is typical of the solar neighborhood and overall consistent with the metallicity distribution function for the Milky Way thin disk in the solar neighborhood; for comparison we use metallicities from the APOGEE Survey \citep{majewski2017} for stars with $|z|$ $<$0.5 kpc and having Galactocentric distances, 7 kpc $\lesssim$ R$_{g}$ $\lesssim$ 9 kpc \citep{hayden2015} to define the local Galactic metallicities and this is shown as the solid black line in Figure \ref{fig:feh_hist}; 
the APOGEE metallicity distribution is also overall consistent with the metallicity distribution obtained for a sample of solar neighborhood stars from the GALAH survey \citep{hayden2020}.
Further statistical comparisons using the APOGEE metallicity distribution as a fiducial distribution will be discussed in Section \ref{sec:feh_rpl}.
The results used in \cite{hayden2015} are from a previous APOGEE data release, which are not on the metallicity scale of APOGEE DR16 (the latter are compared to our results in Figure \ref{fig:feh_comp}, see discussion in Section \ref{sec:feh_comp}). The systematic differences between the \cite{hayden2015} adopted scale and APOGEE DR16 are such that there is no metallicity correction needed to compare with the metallicities in this study.

\subsection{Comparisons with Metallicities from the Literature}
\label{sec:feh_comp}

\begin{figure*}
\gridline{\hspace{-1cm}
          \fig{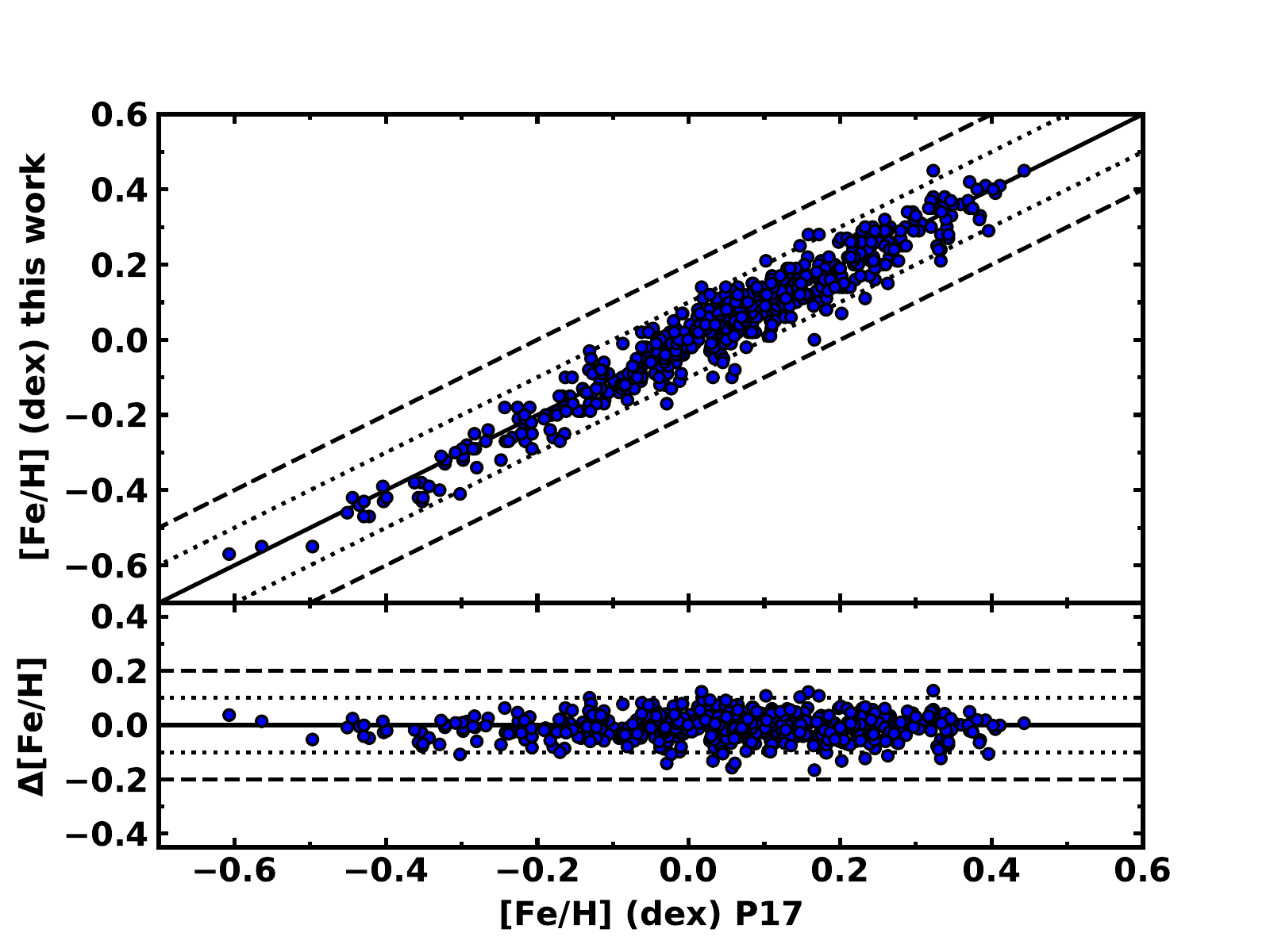}{0.6\textwidth}{}
          \hspace{-0.7cm}
          \fig{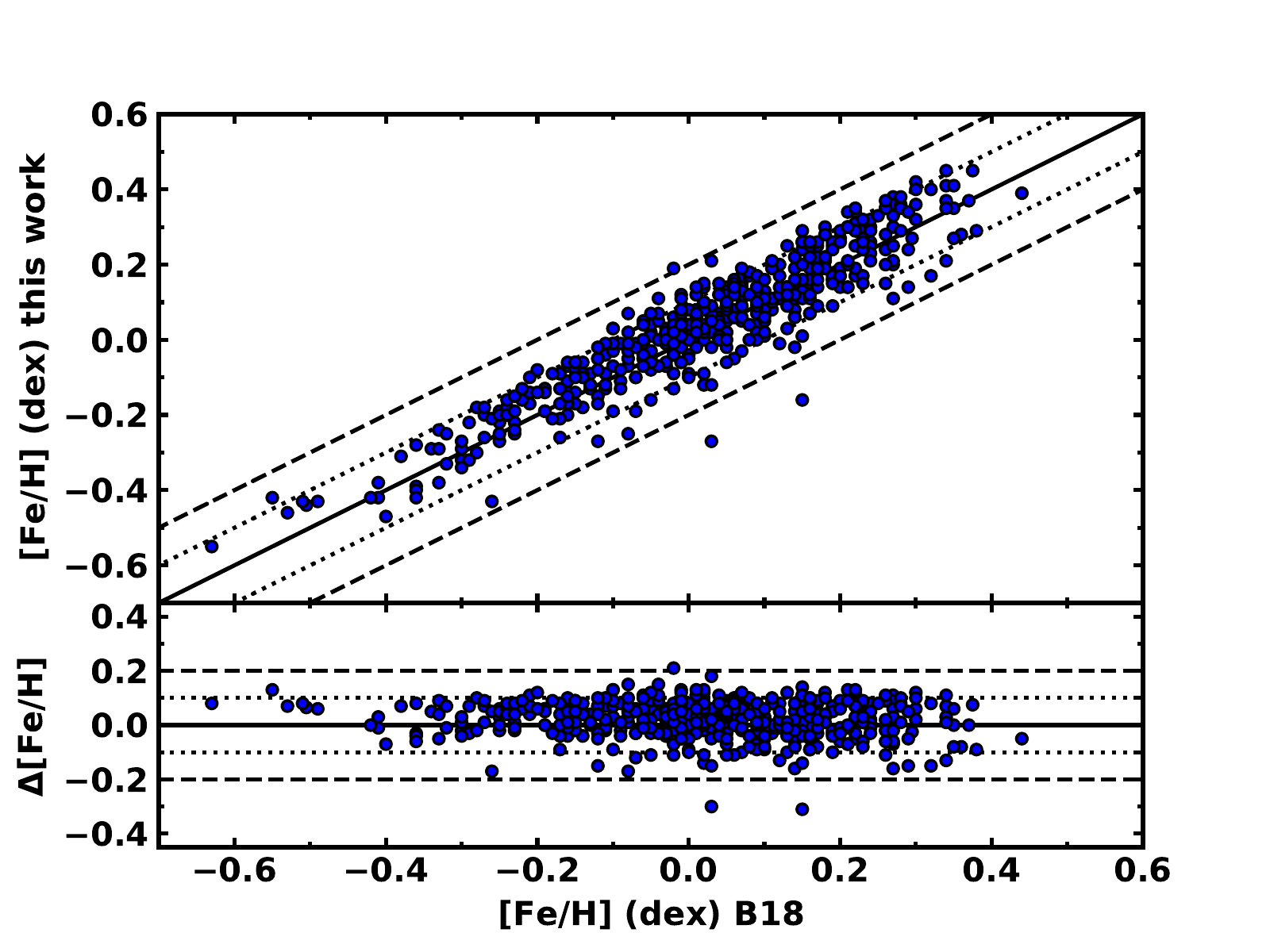}{0.6\textwidth}{}
          }
\vspace{-1.1cm}
\gridline{\hspace{-1cm}
          \fig{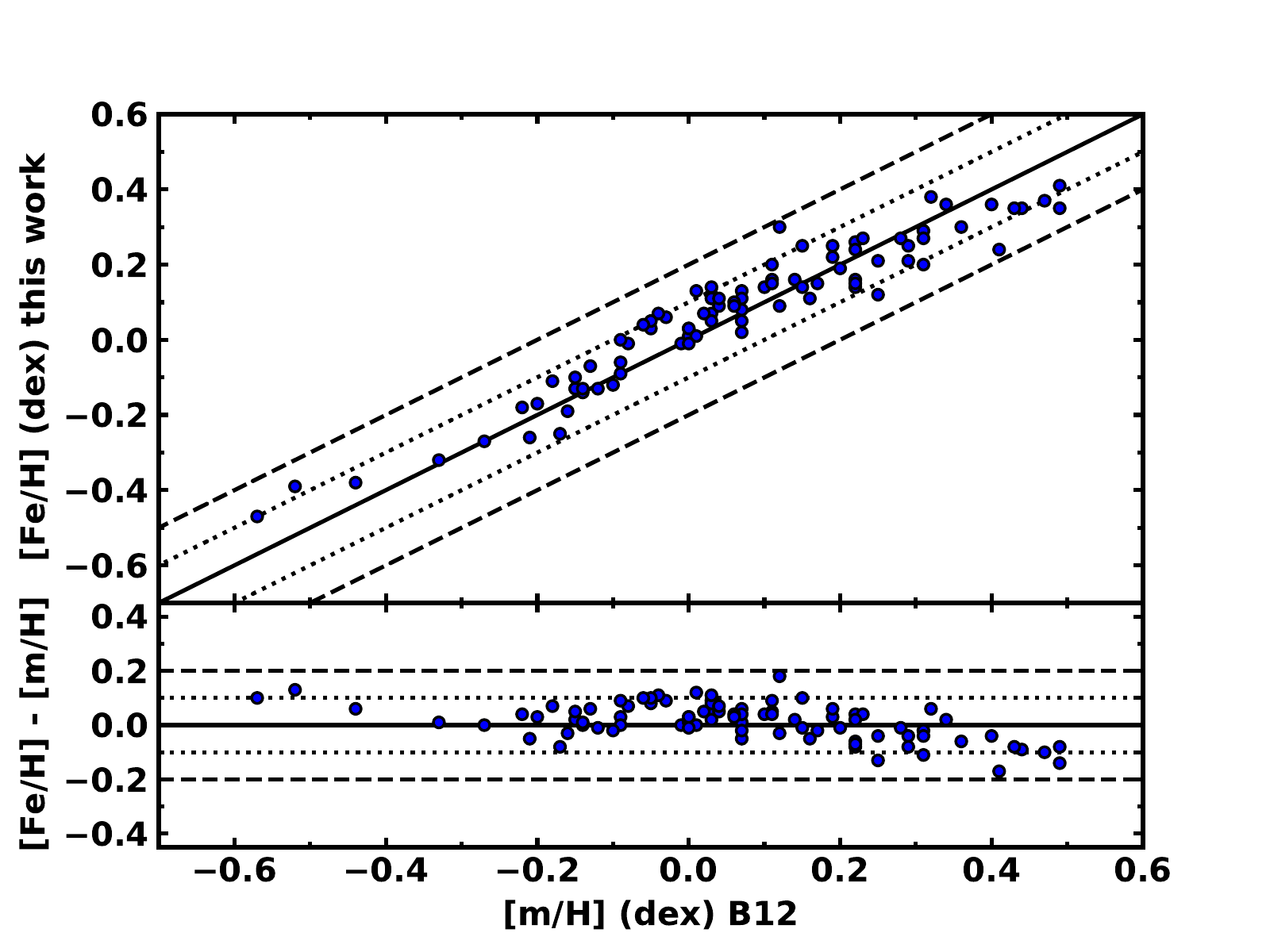}{0.6\textwidth}{}
          \hspace{-0.7cm}
          \fig{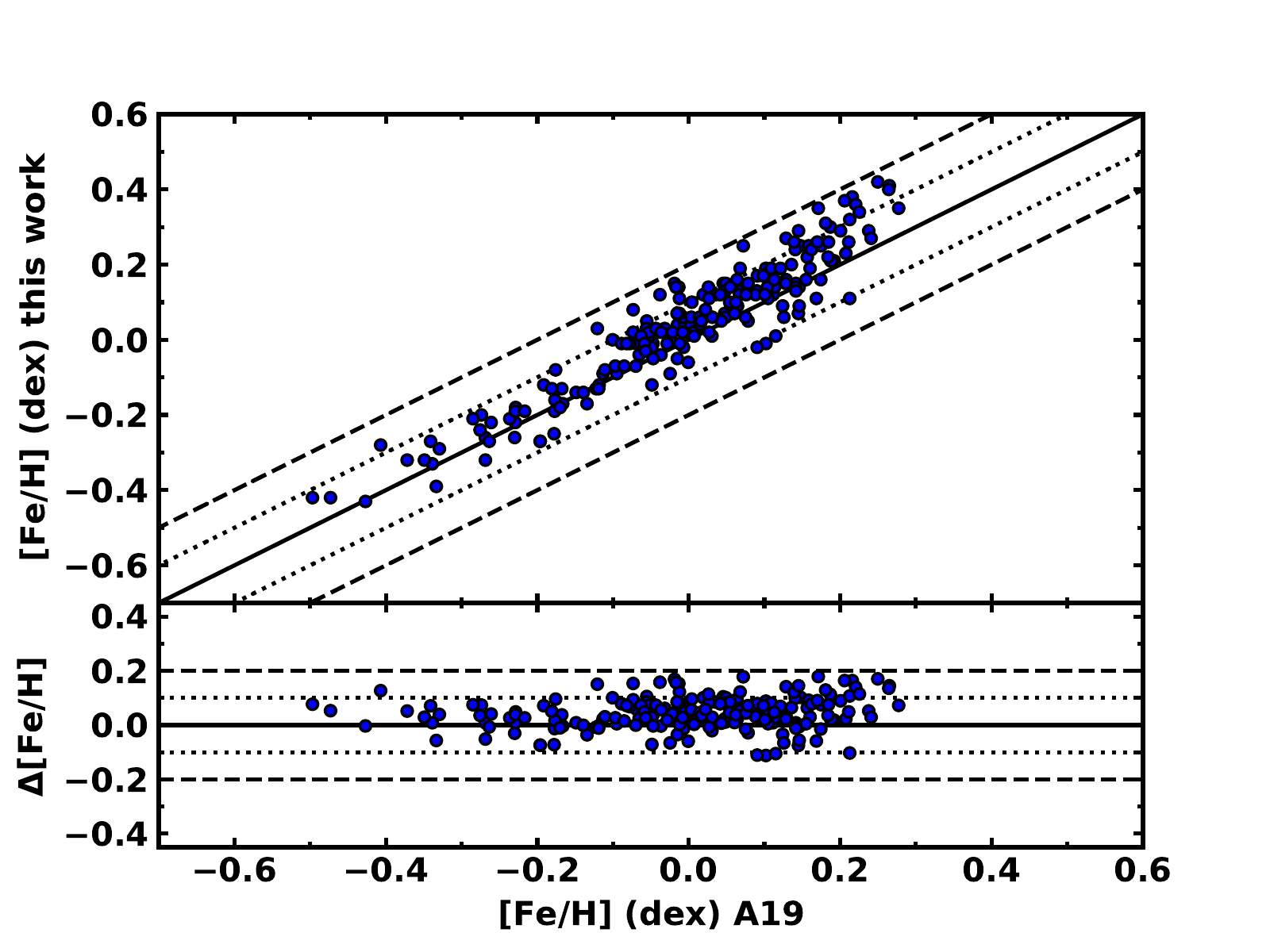}{0.6\textwidth}{}
          }
\vspace{-1.1cm}
\gridline{\fig{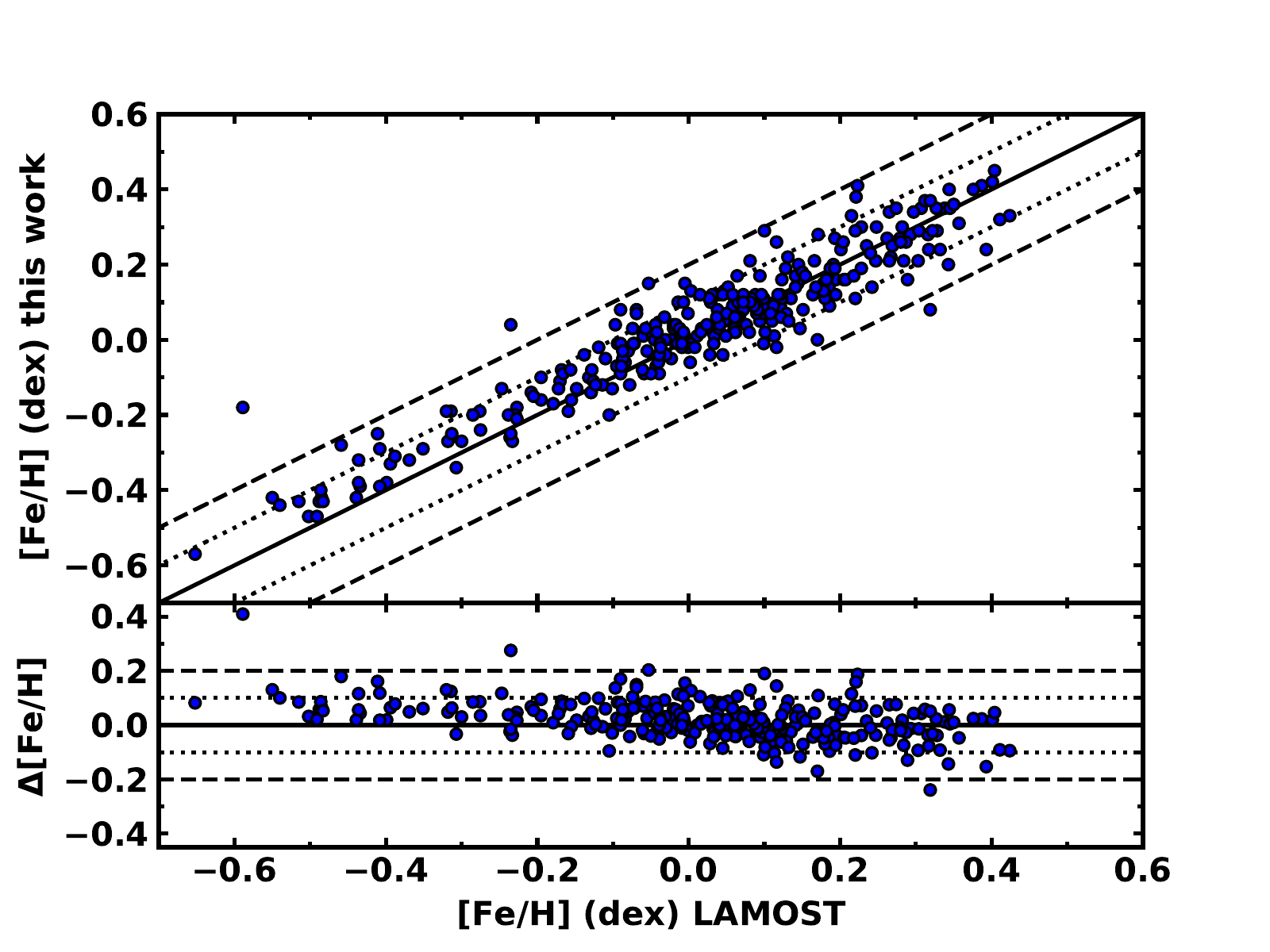}{0.6\textwidth}{}
          }
\vspace{-0.6cm}
\caption{Comparison of metallicities derived in this work with those from \cite{petigura2017a} (P17; upper left panel), \cite{brewer2018} (B18; upper right panel), \cite{buchhave2012} (B12; middle left panel), APOGEE DR16 (A19, middle right panel) and LAMOST DR5 (lower panel). Upper and lower figures within each panel present the direct comparison between the two sets of metallicities and the difference $\Delta$[Fe/H] = [Fe/H] (this work) $-$ [Fe/H] or [m/H] (literature), respectively. Solid lines represent a perfect agreement, while dotted and dashed lines show, as a reference, differences of $\pm$0.1 dex and $\pm$0.2 dex, respectively.}
\label{fig:feh_comp}
\end{figure*}

The CKS sample was first analyzed by \cite{petigura2017a} and more recently by \cite{brewer2018}. Although the details of their methodologies are different, both studies derive [Fe/H] using spectral synthesis instead of the classical spec\-tros\-co\-pic equivalent width method adopted in our study. \cite{petigura2017a} average the parameters obtained from SpecMatch and SME@XSEDE (after putting the latter onto the scale of the former), which is a Python wrapper for the Spectroscopy Made Easy \cite[SME;][]{piskunov2017} software. \cite{brewer2018}, on the other hand, use the results from SME only. 
In Figure \ref{fig:feh_comp}, we present comparisons of our metallicities for the ``clean'' sample with those derived by \cite{petigura2017a} (P17, upper left panel) and \cite{brewer2018} (B18, upper right panel). For 561 stars in common, the distribution of the residuals is Gaussian ($p_{SW} = 0.06$) and the average difference between this work and P17 is -0.01 $\pm$ 0.05 dex.
A linear fit to the comparison between the two sets of metallicities returns a slope 1.00 $\pm$ 0.01 and a Pearson's correlation coefficient of 0.97, which shows an excellent agreement. For B18, there are 493 common targets and the distribution of the residuals is not Gaussian ($p_{SW} \approx 2 \times 10^{-10}$). The median difference is 0.03 $\pm$ 0.04 (MAD) dex and the slope of the linear fit to the comparison between the two sets of metallicities is 0.96 $\pm$ 0.02, with a Pearson's correlation coefficient of 0.93. Thus, the two sets are consistent within 2$\sigma$. We investigated the results for those three stars with most discrepant results: KOI-722, KOI-1311 and KOI-1863. For KOI-722, \cite{brewer2018} give [Fe/H] = 0.15 while our value is [Fe/H] = -0.16. The values in \cite{petigura2017a} and LAMOST DR5 are -0.11 and -0.16, respectively, which are consistent with ours. For KOI-1311, the situation is similar. While \cite{brewer2018} give [Fe/H] = 0.03, our metallicity (-0.27) is consistent with the values from \cite{petigura2017a} and LAMOST DR5 (-0.24 and -0.32, respectively). For KOI-1863, once again, our metallicity (0.19) is different from the one given by \cite{brewer2018} (-0.02), but consistent with the values from \cite{petigura2017a} and LAMOST DR5 (0.20 and 0.23, respectively).


We also show the comparison of our metallicities with those from the earlier analysis of \cite{buchhave2012} in Figure \ref{fig:feh_comp} (B12, middle left panel), which relied on spectral synthesis performed with the Stellar Parameter Classification (SPC) software. The high-resolution spectra were obtained as part of the Kepler Follow-up Program (KFOP). It should be noted, however, that the metallicities provided by \cite{buchhave2012} are [m/H], which are derived over the spectral region 5050 -- 5360 \AA\ via comparisons to library spectra where all elements heavier than helium (``metals'') have their abundances varied over scaled-solar values. 
Our ``clean'' sample has 87 stars in common with B12 and the distribution of the residuals is Gaussian ($p_{SW} = 0.63$), with an average difference of 0.01 $\pm$ 0.07 dex.
The linear fit to the comparison between the two sets of metallicities has a slope 0.84 $\pm$ 0.03 and a Pearson's correlation coefficient of 0.96. These results reveal a significant trend that can also be seen in the residuals, with our metallicities being lower by $\sim$0.1 dex for B12 values of [Fe/H] $\gtrsim$0.2 dex. The reason for this systematic difference is not clear, but it could be caused by differences in the data (e.g., spectral resolution) and adopted methodologies.
We note, however, that similar trends would be observed if we replaced our metallicities with the ones from \cite{petigura2017a} (same 87 stars in common) or \cite{brewer2018} (80 stars in common) and note that systematic differences have also been reported by \cite{furlan2018} when comparing results from SPC code and other software for spectroscopic analyses.

Metallicities for 206 KOIs from our sample are also available from Data Release 16 \citep[DR16;][A19]{ahumada2019} of the Apache Point Observatory Galactic Evolution Experiment \cite[APOGEE;][]{majewski2017} survey. These metallicities were derived automatically from near-infrared (1.5 - 1.7 $\mu$m), high-resolution ($R$ $\sim$ 22,500) APOGEE spectra using the ASPCAP pipeline \citep{garcia2016} and a customized line list \citep{smith2021}. 
The distribution of the residuals is Gaussian ($p_{SW} = 0.04$) with an average difference of 0.04 $\pm$ 0.06 dex. 
The slope of the linear fit to the comparison between the two sets of metallicities (middle right panel of Figure \ref{fig:feh_comp}) is 1.06 $\pm$ 0.03 and the Pearson's correlation coefficient is 0.95, i.e., they are consistent within $2\sigma$. 

Our sample has also 294 KOIs for which metallicities were released as part of Data Release 5 of the lower resolution survey LAMOST (DR5; \citealt{cui2012,ren2016,zong2018})\footnote{http://dr5.lamost.org/}. 
The distribution of the residuals is not Gaussian ($p_{SW} \approx 3 \times 10^{-6}$), the median metallicity difference is 0.02 $\pm$ 0.04 (MAD) dex and the slope of the linear fit to the comparison between the two sets of metallicities is 0.85 $\pm$ 0.02, with a Pearson's correlation coefficient of 0.95.
As previously discussed in the comparison with B12, these results show that there is a significant trend on a global level that can also be noticed in the residuals, with our metallicities being slightly higher by $\sim$0.1 dex for [Fe/H] (LAMOST) $\lesssim$ -0.25 and lower by $\sim$0.05 for [Fe/H] (LAMOST) $>$ +0.3 (see lower panel of Figure \ref{fig:feh_comp}). Such systematic differences are small but metallicity dependant and given the different methodologies, spectral resolution, line lists, atomic data adopted here and in LAMOST, it is not straightforward to identify what could be the reason for discrepancies.
The four stars with the largest differences are: KOI-1050, KOI-1115, KOI-1196 and KOI-2169. For KOI-1050, [Fe/H] in LAMOST's DR5 is -0.24, while our value is 0.04. The value in \cite{petigura2017a} is consistent with ours (0.08). 
For KOI-1115, the metallicities are more scattered: -0.05 in LAMOST's DR5, 0.06 in \cite{brewer2018}, 0.05 in APOGEE's DR16; and our result of 0.15 is in better agreement with the value of 0.12 in \cite{petigura2017a}. For KOI-1196, our metallicity (-0.18) is discrepant from LAMOST's value (-0.59), but much more consistent with other literature mesaurements: -0.16 in \cite{petigura2017a}, -0.24 in \cite{brewer2018} and -0.17 in APOGEE's DR16. For KOI-2169, LAMOST DR5 give [Fe/H] = 0.32, while our value (0.08) is consistent with the measurements from  \cite{brewer2018} (0.10 dex) and in agreement within 0.1 dex with \cite{petigura2017a} who find 0.18 dex.

All in all, the comparisons discussed above indicate good agreement between our metallicities and those from the literature, with median systematic differences typically less than 0.05 dex (that is less than $\sim$0.1 dex, which is typical for external uncertainties in [Fe/H]; see, e.g., \citealt{furlan2018}), and exhibit either no trend (P17), small trends within 1--2$\sigma$ (B18 and A19), or more significant trends (5--7$\sigma$ in the cases of B12 and LAMOST), as measured from the slopes of linear fits to the comparison between the different sets of metallicities.

\section{Discussion}
\label{sec:discussion}

\subsection{Stellar Metallicities and Planetary Radii}
\label{sec:feh_rpl}

We begin to investigate some of the well-explored relations within the sample itself between metallicity of the host star and planetary radii, though not yet taking into consideration relations with planetary orbital periods. 

\subsubsection{Metallicity and the Transition from Small to Large Planets}

\begin{figure}
\plotone{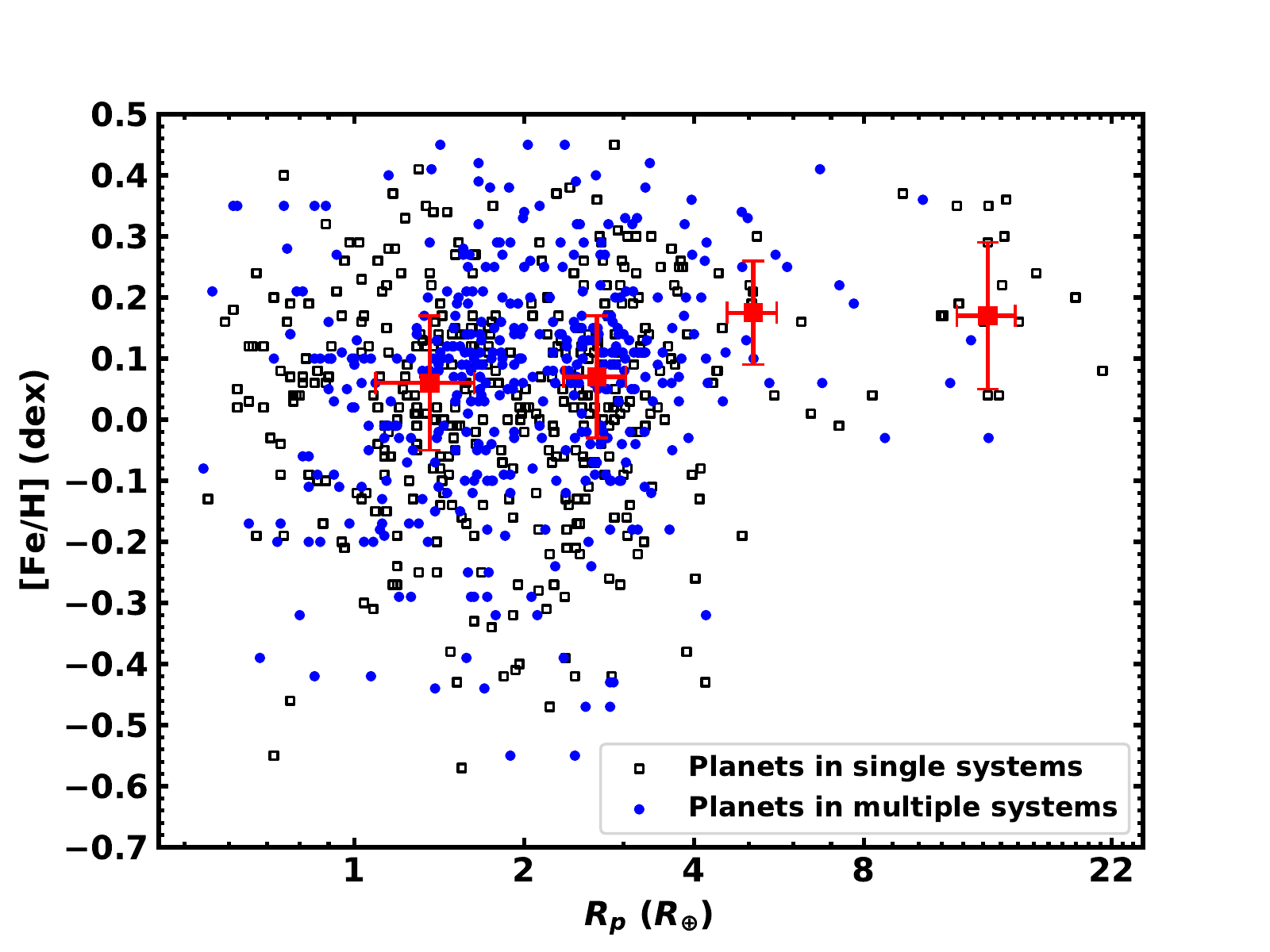}
\caption{Stellar metallicities as a function of the planetary radii for the sample of 804 planets with host star metallicities derived in this study. The planetary radii $R_{p}$ are from Paper I. Black open squares show planets in single systems and filled blue circles represent planets in multiple systems. Red squares show the median metallicities and radii for the following intervals: $R_{p} < 1.9 $ R$_{\oplus}$, $1.9 $ R$_{\oplus} \leq R_{p} < 4.4 $ R$_{\oplus}$, $4.4 $ R$_{\oplus} \leq $ R$_{p} < 8.0 $ R$_{\oplus}$ and $R_{p} \geq 8.0 $ R$_{\oplus}$. The red error bars represent the median absolute deviations around these median values.}
\label{fig:feh_transition}
\end{figure}

As illustrated in Figure \ref{fig:feh_transition}, the distribution of host star [Fe/H] values as a function of $R_{p}$ exhibits a marked change at the boundary between the Sub-Neptune and Sub-Saturn regimes; this transition is observed to occur over a very narrow range in planetary radii. This observation was noted and discussed in \cite{petigura2018}, who quantified the average increase in metallicity ([Fe/H]) across this boundary between small and large planets. Given the relatively large numbers of planets and the accuracy to which both planetary radii (a few percent) and metallicity (typically 0.02 dex) are constrained, it is of interest to further investigate the nature of the [Fe/H] distributions as a function of radius across this transition.


Within this sample of transit-selected systems having planets with orbital periods of less than 100 days, smaller planets exist around stars with iron abundances that encompass most of the range in [Fe/H] found in the disk around the solar neighborhood (7 kpc $<$ $R_{g}$ $<$ 9 kpc; see, e.g., \citealt{hayden2015}), while giant planets with such short orbital periods are found around host stars that inhabit, almost exclusively, the extreme, metal-rich tail of the disk metallicity distribution function (MDF). The relative rarity of such metal-rich stars is one part of the reason for the quite low occurrence rates for hot and warm giant planets, as emphasized by \cite{fulton2017}.

In order to study the influence of [Fe/H] on planetary system architectures (i.e., effect on planetary sizes and orbital periods), the metallicity distributions of host stars were investigated as a function of their daughter planetary radii. 
One method to map quantitatively the changing differences in metallicity distributions as a function of planet radii, is to compare them relative to a comparison, or fiducial, distribution, which is taken here to be a subsample of APOGEE stars that are representative of the local Galactic neighborhood, as defined in Section \ref{sec:feh_dist} and comprising 13105 stars (the MDF for this sample was shown in Figure \ref{fig:feh_hist}).
The APOGEE targets that define the local disk MDF are red giants, the bulk of which have masses of $M_{\star}$ $\sim$ 0.8 -- 1.5 M$_{\odot}$ \citep{pinsonneault2018}, meaning that the majority of the main-sequence progenitors of the APOGEE red giants are FGK stars, which represent most of the CKS sample. The volume of the Galaxy sampled by these APOGEE stars includes the region observed by Kepler, with likely significant overlap in stellar populations.  A simple inspection of Figure \ref{fig:feh_transition} 
reveals that the [Fe/H] distributions of host stars undergoes a striking transition for planets having radii somewhere near $R_{p} \sim$ 4 R$_{\oplus}$; this transition has been found, noted, and discussed in a number of previous studies \citep[e.g.,][]{fulton2017,petigura2018}. We conducted statistical tests in order to better identify and quantify critical planetary radii where the host-star metallicities change significantly. 

The CKS sample that is used in this comparison consists of the ``clean'' sample of 561 stars (Section \ref{sec:feh}). In addition, the 102 stars that were rejected from the ``clean'' sample because they had large errors in their automatically-measured equivalent widths, due to a combination of low-S/N spectra or large projected rotational velocities, are added back to the metallicity sample by adopting the [Fe/H] values from \cite{petigura2017a}. This addition was used to improve statistics and recover the original sample of 663 stars from Paper I. The methods used by \cite{petigura2017a} to derive metallicities involved two codes, which according to that study, can model well and obtain accurate results for low-S/N spectra (SpecMatch), and, in addition, they employ spectrum synthesis (SME@XSEDE) which can handle the rotationally-broadened lines better than our automatic methods for measuring equivalent widths for stars rotating with measurable $v$ sin $i$ values (higher than $\sim$7 km s$^{-1}$; $\sim$ 40 stars in the removed sample have $v$ sin $i$ values above this threshold).
This addition is also justified because the metallicity comparisons described in Section \ref{sec:feh_comp} found that our [Fe/H] values compare very well with the metallicities from \cite{petigura2017a}, with small differences that are Gaussian-distributed and having a mean difference between this work and \cite{petigura2017a} of -0.01$\pm$0.05 dex; median and MAD of -0.01$\pm$0.03 dex; so adding these results will not skew the distribution. We note that in any case, for the 102 removed stars, the systematic difference between our metallicities and \cite{petigura2017a} is just slightly higher, having a median difference of -0.048 $\pm$ 0.053(MAD) dex. 
From here on, when we mention our stellar sample, we refer to the ``clean'' sample of 663 stars unless otherwise stated.

Investigations to define the optimal choices for identifying radii transitions consisted of a number of statistical tests to measure differences in means and variances between the APOGEE local Galactic neighborhood metallicity distribution, compared to those of systems with transiting planets, as a function of planetary radii. The comparison samples were varied by including all systems having their largest known planet up to an initial boundary of R$_{p}$ = 1.9 R$_{\oplus}$, and then expanding this boundary up to R$_{p}$ = 8.0 R$_{\oplus}$, in steps of 0.1 R$_{\oplus}$. At each step, the sample was compared to the APOGEE reference sample. We note that boundary at $R_{p}$ = 1.9 R$_{\oplus}$ was taken from the position of the radius gap as determined in Paper I. The boundary at $R_{p}$ = 8.0 R$_{\oplus}$ was adopted based on the physical arguments discussed by \cite{petigura2018}.

We first employed the Shapiro-Wilk test for normality on the metallicity distribution for planetary systems to inform our choice of statistical tests. We rejected the null hypothesis that the iron abundances were drawn from a normal distribution at each transition radius tested (typically, $p\sim 1 \times 10^{-2} - 2 \times 10^{-6}$). Thus, four different tests were carried out, all being two-sample non-parametric: Mann-Whitney U test to investigate the equality of medians; Levene test to check for equality of variances, replacing the mean as the measurement of central tendency with the median, also known as the Brown-Forsythe test; Cucconi test to jointly test for differences in the central tendency and variance of two independent samples \footnote{We used the Python implementation developed by Grzegorz Mika and available at https://github.com/GrzegorzMika/NonParStat. Unless otherwise stated, we chose the option ``bootstrap'' for the method and 10$^{5}$ replications.}; and Kolmogorov-Smirnov test for completeness, although it is not considered as sensitive as the previous tests. 

\begin{figure*}
\gridline{\fig{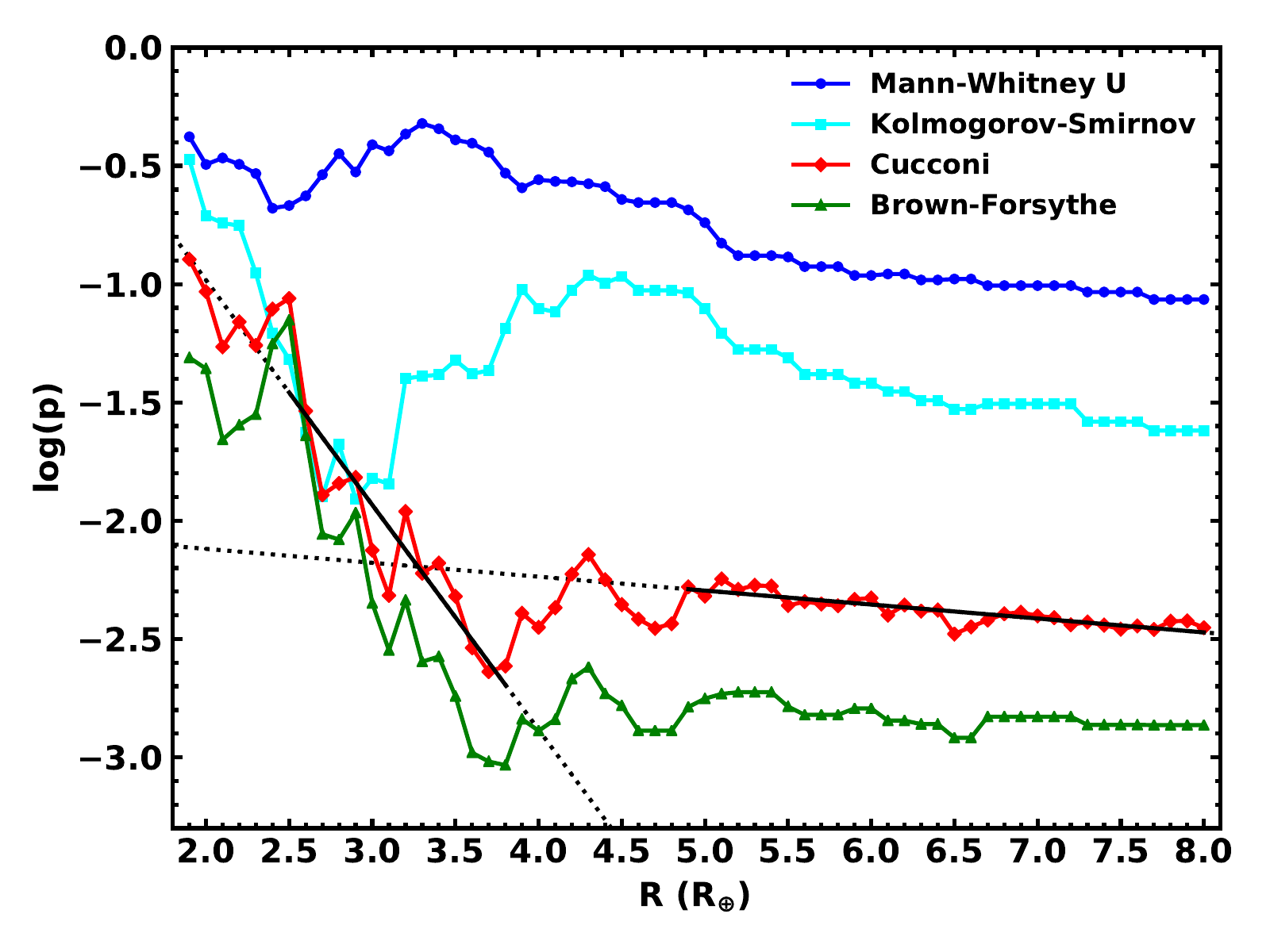}{0.8\textwidth}{}
          }
\gridline{\fig{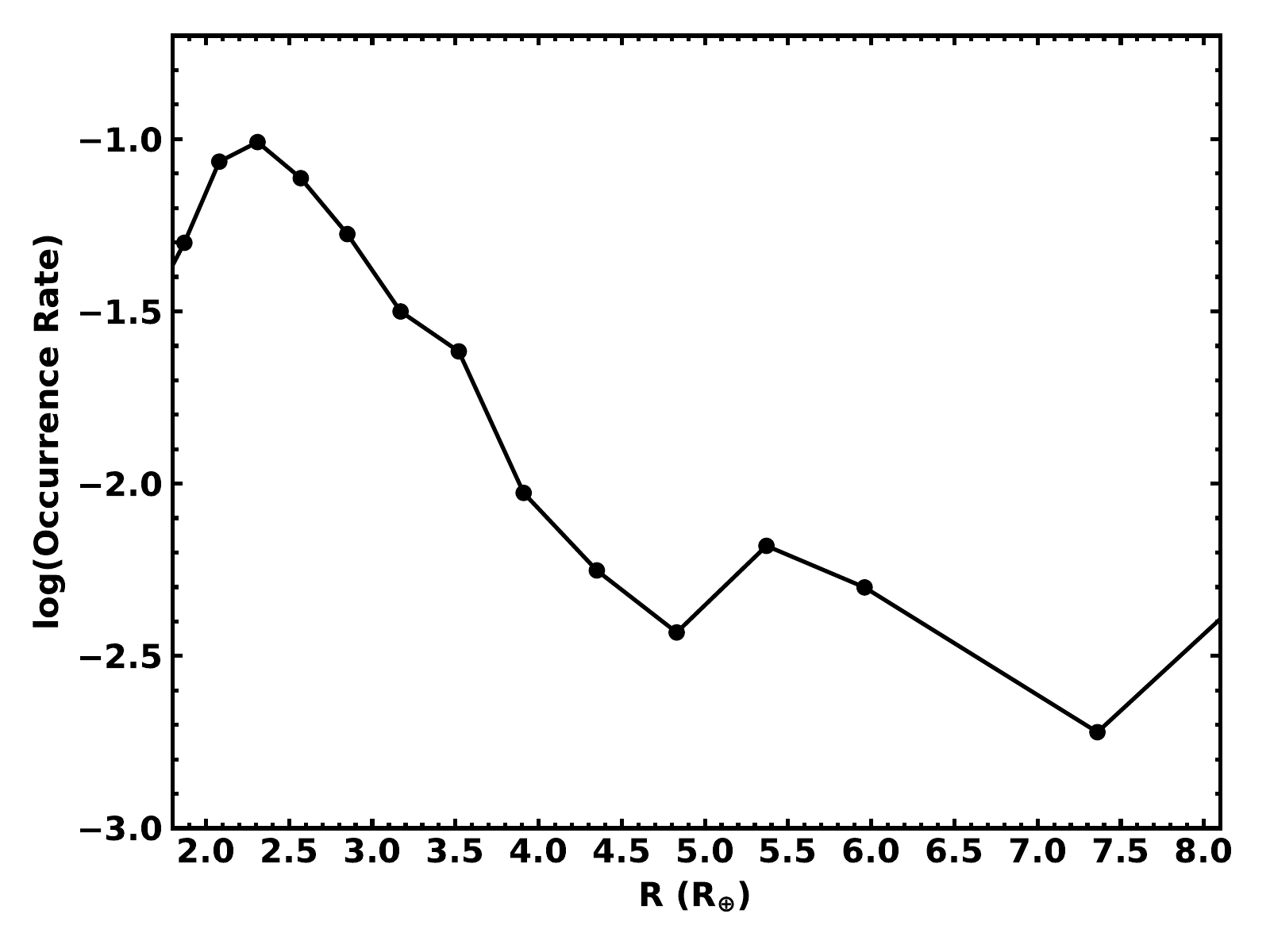}{0.8\textwidth}{}
          }
\caption{\textit{Top panel:} The logarithm of the $p$-values of the statistical tests performed as a function of the planetary radius chosen to represent the transition between systems with small and large planets. Results from the Mann-Whitney, Kolmogorov-Smirnov, Cucconi and Brown-Forsythe tests are represented by blue circles, cyan squares, red diamonds and green triangles, respectively. The black solids lines represent the independent fits performed to the points in the intervals 2.5 - 3.8 $R_{\oplus}$ and 4.9 - 8.0 $R_{\oplus}$. Black dotted lines show the extrapolation of these fits outside the respective intervals. \textit{Bottom panel:} The logarithm of the occurrence rates from \cite{fulton2017} as a function of the mean planetary radius of the corresponding interval from their Table 3.}
\label{fig:radius_cut}
\end{figure*}

The results of these tests are presented in the top panel of Figure \ref{fig:radius_cut} as values of $\log(p)$ versus planetary radius, where the radius represents the upper limit to the systems included in the metallicity distributions. 
For instance, for a transition radius of 3.0 R$_{\oplus}$, a system with a single planet with radius 2.5 R$_{\oplus}$ or two planets with radii 2.0 and 2.5 R$_{\oplus}$ is included in the sample, while a system with a single planet with radius 3.5 R$_{\oplus}$ or with two planets with radii 2.0 and 3.5 R$_{\oplus}$ is not.
The K-S and Mann-Whitney U tests do not return p-values lower than 0.01 for any transition radius, meaning that the metallicity distributions and medians, respectively, are not significantly different from the solar neighborhood stars. On the other hand, the Brown-Forsythe and Cucconi statistical tests yield rather similar results in which the systems with only small planets ($R_{p} < $ 1.9 - 3.0 $R_{\oplus}$) orbit stars that have [Fe/H] distributions that do not differ significantly from the Galactic disk stars ($\log(p)\sim$ -1.0 to -2.0). 
As the planetary sample expands to include ever-larger planets, the [Fe/H] distributions become increasingly different from that of the Galactic background over the regime spanned by Sub-Neptune planets. The abrupt drop in the Cucconi p-values reveals the strong dependence on [Fe/H] for the occurrence rates of Sub-Neptunes, as identified by both \cite{petigura2018} and \cite{owen2018}. These studies suggested that the strong metallicity dependence for the formation of Sub-Neptunes might be due to the enhanced cooling rates in planetary atmospheres caused by high-metallicity, which moderates the planetary atmospheric mass-loss rates.


The differences in the [Fe/H] distributions between the two samples become significant for stars hosting only planets smaller than about $R_{p} \sim $ 2.7 - 3.0 $R_{\oplus}$. The slope defining the rapid decrease in $\log(p)$ values above $R_{p}$ = 2.5 $R_{\oplus}$ (-0.95 $\pm$ 0.12) changes abruptly at $R_{p}$ = 3.9 $R_{\oplus}$, where a transition occurs leading to a different power law for samples with planets having $R_{p} \leq$ 4.9 $R_{\oplus}$ (with a slope -0.06 $\pm$ 0.01). 
In both tests, the transitional behavior in the range 3.9 -- 4.9 $R_{\oplus}$ is similar and exhibits an inflection point at $R_{p}$ $\sim$ 4.3 $R_{\oplus}$.
We adopt $R_{p}$ = 4.4 $R_{\oplus}$ (the middle of the transition range of 3.9 -- 4.9 $R_{\oplus}$) as the transition radius between systems containing only small (Super-Earths and Sub-Neptunes) and large (Sub-Saturns and Jupiters) planets, with an uncertainty of about $\pm$0.5 $R_{\oplus}$.
We note again that the planets included in these tests arise from transiting systems and have orbital periods of less than 100 days.

The bottom panel of Figure \ref{fig:radius_cut} plots occurrence rates of planets of different radii taken from the \cite{fulton2017} analysis of the CKS sample (their Table 3). The occurrence rates derived by \cite{fulton2017} also behave as a broken power-law as a function of planetary radius, with a possible break near $\sim$4.4 $R_{\oplus}$, very similar to the results derived when comparing their host-star metallicities to the Galactic disk [Fe/H] distribution. This comparison indicates that stellar metallicity plays a significant role in defining the size distribution of planets, beginning with the Sub-Neptune size regime, confirming the results derived by \cite{petigura2018}, who utilized different analysis strategies. 

\subsubsection{The Influence of Metallicity on the Planetary System Architecture}
\label{feh_arch}

\begin{figure}
\plotone{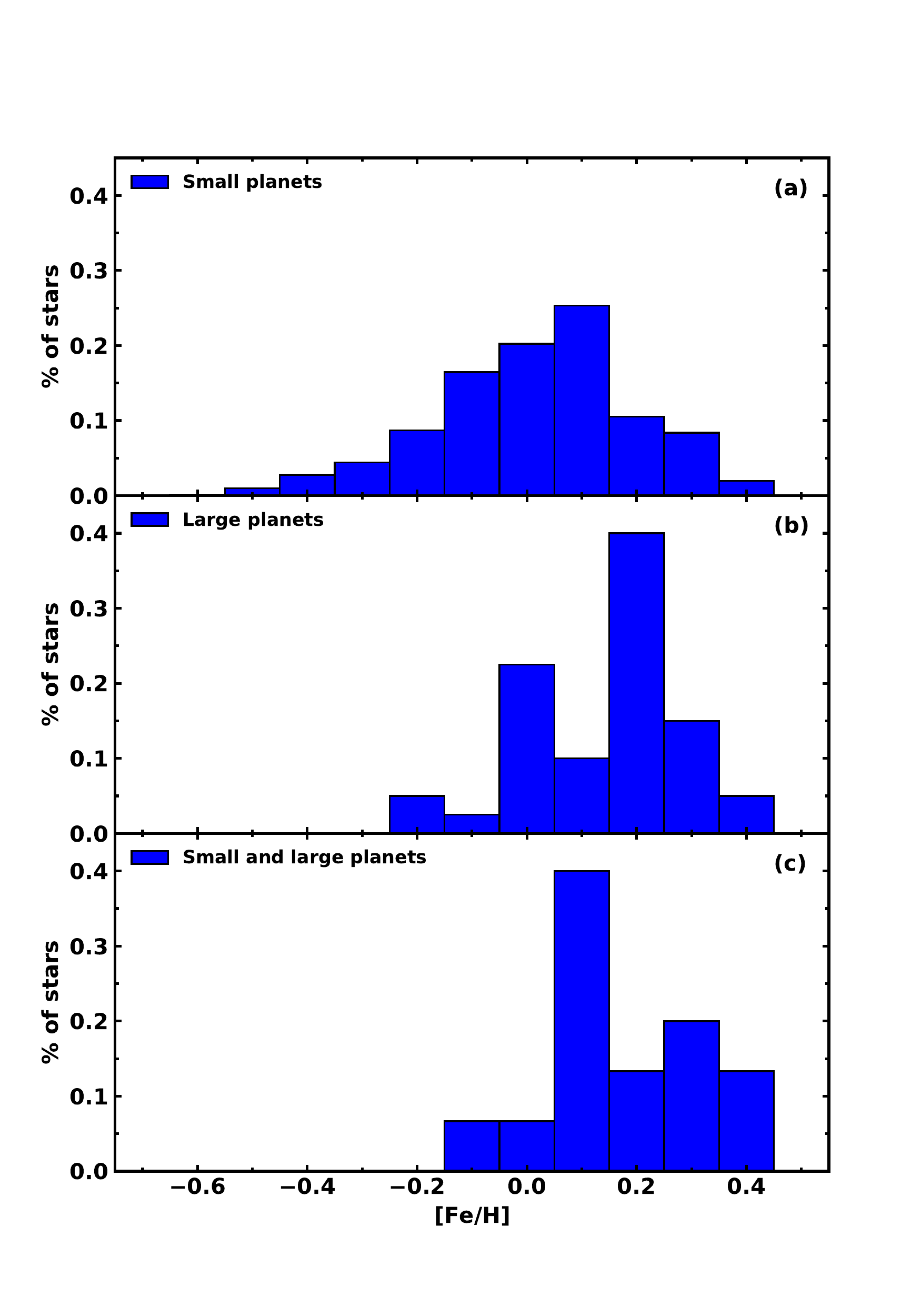}
\caption{Metallicity distributions for systems having exclusively small ($< 4.4$ R$_{\oplus}$) planets (panel a), exclusively large ($\geq 4.4$ R$_{\oplus}$) planets (panel b), and systems having both small and large planets (panel c).}
\label{fig:feh_hist_systems}
\end{figure}

We now subdivide the CKS sample of stars into categories according to their planetary systems and show their metallicity distributions in Figure \ref{fig:feh_hist_systems}. 
Systems having exclusively small ($< 4.4$ R$_{\oplus}$; i.e., single or multiple transiting Super-Earths - SE - or Sub-Neptunes - SN) planets detected are shown in panel (a), systems having exclusively large ($\geq 4.4 R_{\oplus}$; i.e., single or multiple transiting Sub-Saturns - SS - or Jupiters - JP) planets in panel (b), and systems having both small and large planets are shown in panel (c).
It should be noted that the metallicity distributions are stellar metallicities, such that each planetary system appears once in the histograms.


An examination of panel (a) in Figure \ref{fig:feh_hist_systems}, which shows the metallicity distribution of systems having only small planets (Super-Earth and Sub-Neptunes with $R_{p} \leq 4.4 R_{\oplus}$), reveals clearly that the host stars cover effectively the full range in metallicities between roughly -0.6 and +0.4 dex, a result that has been found in previous studies \citep[e.g.,][]{buchhave2012,buchhave2014,petigura2018}. By segregating the sample into systems that only have small transiting planets detected, it is clear that, even without having closely-orbiting large planets, small planets can form around the most metal-rich stars in the Galaxy ([Fe/H] $\sim$ 0.4 - 0.5), though it is worth noting that the bulk of the metallicity distribution is centered around the solar value, with 
a median metallicity of 0.04 $\pm$ 0.11 (MAD) dex. In this discussion, the samples of Super-Earth and Sub-Neptune hosts were combined, but if treated separately, the medians and MADs of their metallicity distributions are similar (as it will be discussed below).

The metallicity distributions for CKS sample stars hosting only large planets (Figure \ref{fig:feh_hist_systems}, panel b) are clearly concentrated towards higher metallicity values, as found and discussed in several previous studies \citep[e.g.,][]{sousa2008,ghezzi2010a,petigura2018,adibekyan2019}. Their distributions have lower metallicity spreads that appear to exhibit a cut-off at [Fe/H] $\sim$ $-0.2$, not reaching the lowest metallicities in the CKS sample; 
the median metallicity is 0.17 $\pm$ 0.10 (MAD) dex. The MDF for stars only with large planets is quite distinct from that of the local solar neighborhood.

Figure \ref{fig:feh_hist_systems} also shows the metallicity distributions of host stars having large plus small planets (panel c). Host stars having both small and large known planets in our sample are, on average, more metal rich than with small planets only and do not reach metallicities lower than [Fe/H] $\sim$ -0.1.
The median metallicity is 0.13 $\pm$ 0.09 (MAD) dex and, using Cucconi tests, we are able to reject the null hypothesis that they are drawn from the same parent population as systems having only small planets ($p \approx 0.004$). The same result is not true when comparing the small plus large planet systems to systems with only large planets ($p \approx 0.81$).

\subsubsection{Systems with Single and Multiple Planets}
\label{singles_multiples}

\begin{figure}
\plotone{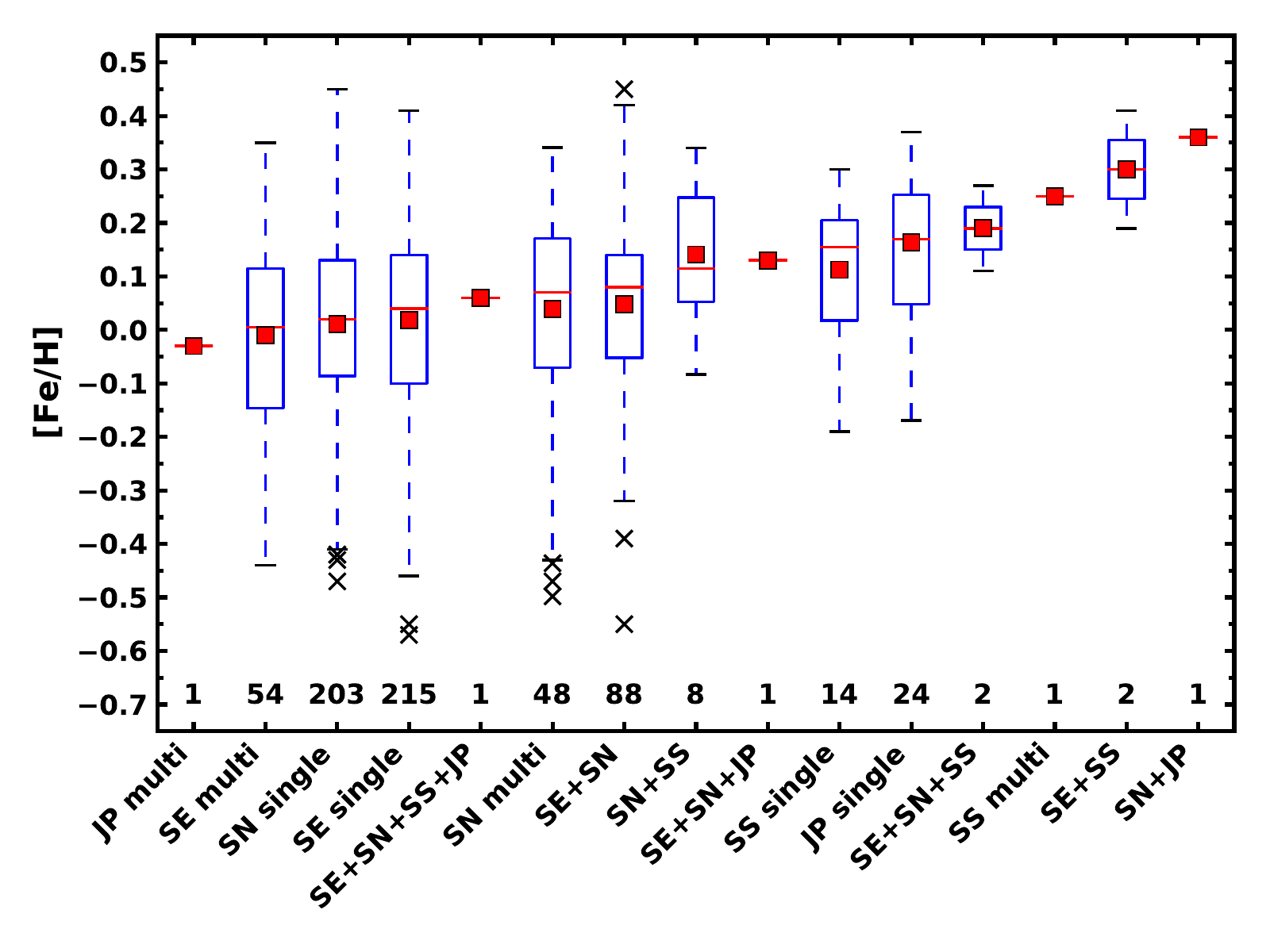}
\caption{Box plots for the metallicity distributions of systems segregated into classes according to Table \ref{tab:pl_feh}. 
Medians and means are represented by red lines and squares, respectively. The blue boxes extend from the 25th (Q1) to the 75th (Q3) percentile, while the whiskers represented by blue dashed lines cover the range between Q1 - 1.5IQR and Q3 + 1.5IQR, where IQR is the inner quartile range. Black crosses show the outliers. Systems are organized in order of ascending median metallicity. The number of systems in each group is shown by the numbers in the lower part of the plot. Groups with just one system are represented by its metallicity value and have no associated statistics.}
\label{fig:feh_systems_box}
\end{figure}

We now subdivide the CKS sample of stars into additional categories, as defined in Table \ref{tab:pl_feh}. Here we use SE single, SN single, SS single and JP single to refer to host stars having a single planet detected, either one Super-Earth ($R_{p} < 1.9 R_{\oplus}$), one Sub-Neptune ($1.9 R_{\oplus} \leq R_{p} < 4.4 R_{\oplus}$), one Sub-Saturn ($4.4 R_{\oplus} \leq R_{p} < 8.0 R_{\oplus}$) or one Jupiter ($R_{p} \geq 1.9 R_{\oplus}$), respectively. The names SE multi, SN multi, SS multi and JP multi correspond to systems having exclusively multiple planets of the same type, while systems with combinations of planet types (e.g., SE+SN, SE+SS, SE+SN+SS+JP) have at least one planet of each kind. In Figure \ref{fig:feh_systems_box} we show the box plot for the metallicities sorted in order of increasing median metallicity, which is represented by the red lines. The number of systems in each one of the categories is shown in the bottom of the figure. Note that in some instances the studied CKS sample has less than three systems representative of a class, while most of the {\it Kepler} host stars contain at least one Super-Earth or Sub-Neptune. The cases with less than three representative systems will not be discussed below but are still shown in Figure \ref{fig:feh_systems_box} for illustrative purposes.


\begin{deluxetable*}{lrrrrcrr}[t!]
\tablecaption{Planetary systems and their median metallicities.
\label{tab:pl_feh}}
\tablecolumns{8}
\tablehead{
\colhead{System} & \colhead{SE} & \colhead{SN} & \colhead{SS} & \colhead{JP} & \colhead{Number of} & \colhead{Median} & \colhead{MAD} \\
\colhead{type} & \nocolhead{} & \nocolhead{} & \nocolhead{} & \nocolhead{} & \colhead{systems} & \colhead{(dex)} & \colhead{(dex)}
}
\startdata
SE single & 1 & 0 & 0 & 0 & 215 & 0.04 & 0.12 \\
SN single & 0 & 1 & 0 & 0 & 203 & 0.02 & 0.11 \\
SS single & 0 & 0 & 1 & 0 & 14 & 0.15 & 0.08 \\
JP single & 0 & 0 & 0 & 1 & 24 & 0.17 & 0.12 \\
SE multi & $>$1 & 0 & 0 & 0 & 54 & 0.01 & 0.14 \\
SN multi & 0 & $>$1 & 0 & 0 & 48 & 0.07 & 0.12 \\
SS multi & 0 & 0 & $>$1 & 0 &  1 & 0.25 & \nodata \\
JP multi & 0 & 0 & 0 & $>$1 &  1 & -0.03 & \nodata \\
SE+SN & $\geq$1 & $\geq$1 & 0 & 0 & 88 & 0.08 & 0.09 \\
SE+SS & $\geq$1 & 0 & $\geq$1 & 0 &  2 & 0.30 & 0.11 \\
SE+JP & $\geq$1 & 0 & 0 & $\geq$1 &  0 & \nodata & \nodata \\
SN+SS & 0 & $\geq$1 & $\geq$1 & 0 &  8 & 0.12 & 0.10 \\
SN+JP & 0 & $\geq$1 & 0 & $\geq$1 &  1 & 0.36 & \nodata \\
SS+JP & 0 & 0 & $\geq$1 & $\geq$1 &  0 & \nodata & \nodata \\
SE+SN+SS & $\geq$1 & $\geq$1 & $\geq$1 & 0 &  2 & 0.19 & 0.08 \\
SE+SN+JP & $\geq$1 & $\geq$1 & 0 & $\geq$1 &  1 & 0.13 & \nodata \\
SE+SS+JP & $\geq$1 & 0 & $\geq$1 & $\geq$1 & 0 & \nodata & \nodata \\
SN+SS+JP & 0 & $\geq$1 & $\geq$1 & $\geq$1 & 0 & \nodata & \nodata \\
SE+SN+SS+JP & $\geq$1 & $\geq$1 & $\geq$1 & $\geq$1 & 1 & 0.06 & \nodata \\
\enddata
\tablecomments{Systems named as ``single'' and ``multi'' have a single (1) or multiple ($>$1) known planets of the specified class, respectively. The expression $\geq$1 indicates the system has at least one planet of the specified class.}
\end{deluxetable*}



The overall trend seen in Figure \ref{fig:feh_systems_box} highlights that systems having larger planets are typically more metal rich when compared to systems having small planets. On the left side of the figure, we find systems containing only planets smaller than Saturn and Jupiter (SE multi, SN single, SE single, SN multi and SE+SN). Their [Fe/H] distributions span wider intervals with a more frequent occurrence of low metallicity ``outliers'' and a median metallicity value that is close to solar ($\lesssim$ 0.1 dex). 

The majority of the systems (456 out of a total of 663, or 69\%) in the studied CKS sample have only one transiting planet detected. For single systems the apparent trend is: SE single and SN single have basically indistinguishable median host star metallicities that scatter roughly around solar, while SS single and JP single orbit host stars with higher median metallicities (see Figure \ref{fig:feh_systems_box} and Table \ref{tab:pl_feh}).
The second most numerous category in the studied sample corresponds to multiple systems composed of only small planets, or, SE multi, SN multi and SE+SN (a total of 190 systems or $\sim$29\% of the sample); their metallicities also have overall lower median values 
than those of systems with larger planets. Based on Cucconi tests for classes with three or more systems (Table \ref{tab:stats_tests_systems}), we do not see significant differences between the metallicity distributions of stars hosting single or multiple Super-Earths or Sub-Neptunes \citep[see also, e.g.,][]{munozromero2018,weiss2018}. In the next Section, however, we will discuss metallicity signatures that appear when we segregate such systems according to their orbital periods.


\startlongtable
\begin{deluxetable*}{lcc}
\tablecaption{Results from Cucconi tests for the samples in Table \ref{tab:pl_feh} with three of more representative systems.
\label{tab:stats_tests_systems}}
\tablecolumns{3}
\tablehead{
\colhead{Samples} & \colhead{Cucconi $p$-value} & Significantly Different?\tablenotemark{a}}
\startdata
SE single -- SN single & 0.8904 & N \\
SE single -- SS single & 0.1052 & N \\
SE single -- JP single & 0.0003 & Y \\
SE single -- SE multi & 0.3446 & N \\
SE single -- SN multi & 0.2020 & N \\
SE single -- SE+SN & 0.2934 & N \\
SE single -- SN+SS & 0.1879 & N \\
SN single -- SS single & 0.08438 & N \\
SN single -- JP single & 0.0002  & Y \\
SN single -- SE multi & 0.5311 & N \\
SN single -- SN multi & 0.1058 & N \\
SN single -- SE+SN & 0.2268 & N \\
SN single -- SN+SS & 0.1289 & N \\
SS single -- JP single & 0.5313 & N \\
SS single -- SE multi & 0.06911 & N \\
SS single -- SN multi & 0.3494 & N \\
SS single -- SE+SN & 0.2176 & N \\
SS single -- SN+SS & 0.7880 & N \\
JP single -- SE multi & 0.0007 & Y \\
JP single -- SN multi & 0.0496 & N \\
JP single -- SE+SN & 0.0031 & Y \\
JP single -- SN+SS & 0.8867 & N  \\
SE multi -- SN multi & 0.2613  & N \\
SE multi -- SE+SN & 0.0653 & N  \\
SE multi -- SN+SS & 0.1059 & N  \\
SN multi -- SE+SN & 0.3369 & N \\
SN multi -- SN+SS & 0.5966 & N  \\
SE+SN -- SN+SS & 0.4603 & N \\
\enddata
\tablenotetext{a}{Y and N indicate if the [Fe/H] distributions are (p-value $\leq$ 0.01) or are not (p-value $>$ 0.01) significantly different, respectively.}
\end{deluxetable*}


As the median host star metallicity increases (towards the right side of Figure \ref{fig:feh_systems_box}), we see systems with at least one large planet (SN+SS, 
SS single, JP single, plus several systems for which we do not have enough statistics but contain a large planet).
The metallicity distributions are narrower and shifted towards higher values of [Fe/H]. The median metallicities in this regime are all $\gtrsim$ 0.10 dex and there are no low metallicity ``outliers''. 
It is notable that there are only a few systems having combinations of large planets or having a large planet plus combinations of small planets when compared to systems with single planets, or with multiple small planets (SE and SN). 

We performed Cucconi tests with the groups in Figure \ref{fig:feh_systems_box} that have three or more systems and the results are presented in Table \ref{tab:stats_tests_systems}. 
Significant metallicity differences are indicated by the smallest p-values ($\leq$ 0.01), which are obtained from certain small planet systems when compared to single Jupiter systems (JP single): SE single ($p_{CC} \approx 3 \times 10^{-4}$), SN single ($p_{CC} \approx 2 \times 10^{-4}$), SE multi ($p_{CC} \approx 7 \times 10^{-4}$), and SE+SN ($p_{CC} \approx 3 \times 10^{-3}$). We find that stars hosting single or multiple Super-Earths (SE single and SE multi, respectively), a single Sub-Neptune (SN single) or at least one Super-Earth and one Sub-Neptune (SE+SN) are not drawn from the same parent population as those hosting a single Jupiter (JP single). 
For all other comparisons, $p > 0.01$, and thus we are not able to distinguish, via metallicity, any of the additional planetary classes shown in Figure \ref{fig:feh_systems_box} based on p-values of Cucconi tests.

\subsection{Stellar Metallicities and Orbital Periods}
\label{sec:feh_per}

\subsubsection{Metallicities of Hot and Warm Systems}

Previous studies have revealed that there is a correlation between host star metallicity and the presence of hot planets with orbital periods less than $\sim$8--10 days \citep[e.g.,][]{mulders2016,petigura2018,wilson2018,dong2018}. Most of these investigations of {\it Kepler} planets considered host star metallicities and the orbital period of each individual planet in a system. Here, we use the derived iron abundances and divide the studied CKS planet sample according to their orbital periods into hot ($P$ $\leq$ 10 d) and warm (10 d $<$ $P$ $\leq$ 100 d), but we take a slightly different approach to investigate possible metallicity trends. As previously in the paper, we consider planetary architecture, or, more specifically in this case the number and types of planets in a system along with their orbital periods. 

The majority of the planetary systems in the CKS sample studied here is composed of single planets (456 single transiting planets detected from {\it Kepler}). It is important to keep in mind, however, that the number of detected transiting planets does not represent the total number of planets in that system but only those that could be detected within {\it Kepler} constraints. In Table \ref{tab:pl_feh_p} we present the classes corresponding to planet types, multiplicity and orbital period, number of systems in each group, along with the median metallicities of their host stars as well as their MADs.


\begin{deluxetable*}{llcrr}[t!]
\tablecaption{Hot and warm planets and median metallicities of their hosts.
\label{tab:pl_feh_p}}
\tablecolumns{5}
\tablehead{
\colhead{Planets} & \colhead{Orbital} & \colhead{Number of} & \colhead{Median} & \colhead{MAD} \\
\colhead{} & \colhead{Period} & \colhead{systems} & \colhead{(dex)} & \colhead{(dex)}
}
\startdata
SE Single & Warm & 48 & -0.05 & 0.10 \\
SE Single & Hot & 167 & 0.06 & 0.11 \\
SN Single & Warm & 153 & 0.02 & 0.11 \\
SN Single & Hot & 50 & 0.04 & 0.10 \\
SS Single & Warm & 7 & 0.08 & 0.09 \\
SS Single & Hot & 7 & 0.16 & 0.08 \\
JP Single & Warm & 7 & 0.17 & 0.01 \\
JP Single & Hot & 17 & 0.20 & 0.15 \\
SE Multi & Warm & 6 & -0.18 & 0.07 \\
SE Multi & Hot & 25 & 0.10 & 0.11 \\
SE Multi & Hot+Warm & 23 & -0.09 & 0.10 \\
SN Multi & Warm & 20 & 0.06 & 0.11 \\
SN Multi & Hot & 8 & 0.11 & 0.09 \\
SN Multi & Hot+Warm & 20 & 0.13 & 0.13 \\
SS Multi & Hot+Warm & 1 & 0.25 & \nodata \\
JP Multi & Warm & 1 & -0.03 & \nodata \\
SE+SN & Warm & 11 & -0.02 & 0.17 \\
SE+SN & Hot & 21 & 0.11 & 0.09 \\
SE+SN & Hot+Warm & 56 & 0.06 & 0.09 \\
SE+SS & Hot & 1 & 0.41 & \nodata \\
SE+SS & Hot+Warm & 1 & 0.19 & \nodata \\
SN+SS & Warm & 3 & 0.13 & 0.10 \\
SN+SS & Hot & 1 & 0.34 & \nodata \\
SN+SS & Hot+Warm & 4 & 0.08 & 0.08 \\
SN+JP & Warm & 1 & 0.36 & \nodata \\
SE+SN+SS & Hot+Warm & 2 & 0.19 & 0.08 \\
SE+SN+JP & Hot+Warm & 1 & 0.13 & \nodata \\
SE+SN+SS+JP & Hot+Warm & 1 & 0.06 & \nodata \\
\enddata
\tablecomments{SE = Super-Earths, SN = Sub-Neptunes, SS = Sub-Saturns, JP = Jupiters. Hot and warm planets have $P$ $\leq$ 10 d and 10 d $<$ $P$ $\leq$ 100 d, respectively.}
\end{deluxetable*}

\begin{figure}
\plotone{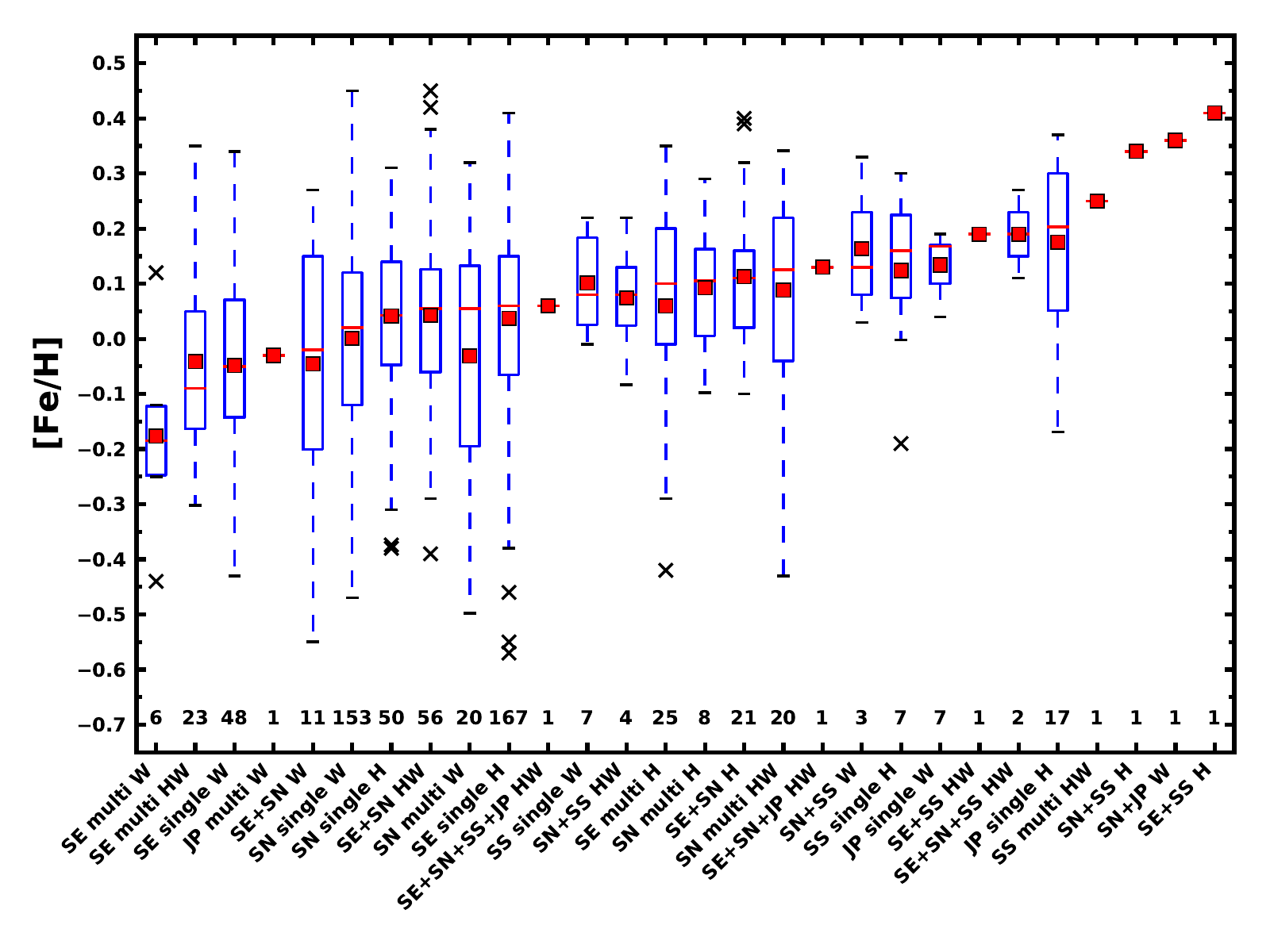}
\caption{Box plots for the metallicity distributions of systems segregated into classes according to planet size (SE, SN, SS, JP), multiplicity (single/multiple) and orbital period (warm W, hot H, hot+warm HW). The symbols are the same as in Figure \ref{fig:feh_systems_box}.}
\label{fig:feh_system_hot_warm}
\end{figure}

Figure \ref{fig:feh_system_hot_warm} shows the median metallicities of host stars (red lines in the box plot diagram) segregated into 20 classes (having 3 or more systems) depending on the types of planets, multiplicity and orbital periods (only hot, only warm and hot+warm combination). The subclasses are ordered based on the median host star metallicity within that subclass, increasing from left to right. The division in subclasses follows the same convention as used previously, e.g., ``SE single H" refers to systems in our sample having only one Super-Earth planet with an orbital period shorter than or equal to 10 d, while ``SE single W'' corresponds to systems with a single Super-Earth with 10 d $<$ $P$ $\leq$ 100 d. Systems marked as HW (e.g., SE+SN HW), for example, have a combination of Super-Earths and Sub-Neptunes with both hot and warm orbital periods. Here we only include in the discussion classes having 3 or more systems.


It is apparent from Figure \ref{fig:feh_system_hot_warm} that as the median metallicity of the host stars increases, the mix of hot (H), warm (W), and hot+warm (HW) subclasses changes. For example, there are four subclasses that have median host star metallicities below solar ([Fe/H] $<$ 0.00) and these contain SE multi and singles, and SE+SN. 
These systems harbor only warm planets and not hot ones, except for ``SE multi HW'', which have both warm and hot Super-Earths.  In particular, 2 out of these 3 subclasses with the lowest median metallicities are composed of multiple and single warm Super-Earths systems. This result is overall consistent with those of \cite{owen2018}, who find that small planets that do not have H-He atmospheres at relatively long periods (Super-Earths W) are more commonly found around low metallicity stars.

One interesting signature that emerges, however, is that the stellar hosts with the lowest median metallicity harbor multiple warm Super-Earths (SE multi W); our sample has six systems and their median metallicity is -0.18 (Table \ref{tab:pl_feh_p}). Except for KOI-2007 with a metallicity [Fe/H] = +0.12 (identified as an outlier X for its box in Figure \ref{fig:feh_system_hot_warm}), all of the SE multi W host stars are metal poor and it is notable that the metallicity distribution of this class does not extend to high [Fe/H], unlike the warm Super-Earths in single systems, which also form at high metallicities. Although based on a small number of systems --- given that Super-Earths at relatively long orbital periods are hard to detect --- this signature may indicate that systems having multiple transiting warm Super-Earths tend not to form at all metallicities, and their formation might be favored around stars with lower metallicities.

Most of the subclasses with median metallicities above solar ([Fe/H] $>$ 0.00) have at least one hot planet. The exceptions are the single and multiple warm Sub-Neptunes (0.02 and 0.06, respectively), single warm Sub-Saturns (0.08), single warm Jupiters (0.17) and systems containing warm Sub-Neptunes and Sub-Saturns (SN+SS W; 0.13). 
At the very metal-rich regime ([Fe/H] $\geq$ 0.25), all four subclasses hosting multiple planets have only one representative and will not be discussed here.

To summarize the trends between system architecture and orbital period classes, we find that: (1) systems with only warm planets consistently have lower median metallicities than systems with any hot planets (H or HW), (2) systems with hot planets of any kind have typically super-solar metallicities, (3) systems with only warm Super-Earths have the lowest median metallicities, and (4) systems with at least one large planet (SS or JP) have super-solar host star metallicities.

Finally, it should be noted that when considering all classes of planets just segregating in systems having only warm planets versus system having any hot planets (H or HW) the median metallicities for hot, hot+warm and warm systems are 0.07 $\pm$ 0.11 (MAD) dex, 0.05 $\pm$ 0.11 (MAD) dex and 0.01 $\pm$ 0.13 (MAD) dex, respectively. Based on Cucconi tests, we find that the hot and warm systems do not belong to the same parent population ($p_{CC} \approx 1 \times 10^{-4}$), while when comparing systems hot+warm with hot ($p_{CC} \approx 0.07$), or, hot+warm with warm ($p_{CC} \approx 0.03$) we cannot reject the hypothesis that they are from the same populations.

\subsubsection{Metallicities of Hot and Warm Single versus Multiple Planet Systems}
\label{sec:feh_hot_warm}

The discussion above of the median metallicities of stellar host classes (Figure \ref{fig:feh_system_hot_warm}) takes into consideration the planetary multiplicity (single or multiple) and orbital periods (hot or warm) that can be further examined for patterns or trends between single and multiple systems. One pattern that appears is in the median values of [Fe/H] between hot and warm planet systems for certain stellar host classes (from Table \ref{tab:pl_feh_p}), where the difference $\Delta$[Fe/H](H - W) can be calculated and examined; for example, in single planet systems $\Delta$[Fe/H] (SE single) = 0.11, $\Delta$[Fe/H] (SN single) = 0.02, $\Delta$[Fe/H] (SS single) = 0.08, and $\Delta$[Fe/H] (JP single) = 0.04.

For the single planet systems, the host star metallicities 
do not vary significantly over planet class (small planets versus giant planets) between the hot and warm systems; the median difference for all planet classes in single systems is $\Delta$[Fe/H] (All Planet Types) = 0.06 $\pm$ 0.03 (MAD) dex, where the MAD of 0.03 dex in [Fe/H] is comparable to the abundance uncertainties.
Although the median differences in the stellar Fe abundances between hot and warm single planetary systems are small, they are consistent with hot planets orbiting stars that are slightly more metal-rich than warm planets. The same differences for multiple planet systems (where there are both hot and warm examples) are, on the other hand, significantly larger, with $\Delta$[Fe/H] (SE multi) = 0.29, $\Delta$[Fe/H] (SN multi) = 0.05, $\Delta$[Fe/H] (SE+SN) = 0.13, $\Delta$[Fe/H] (SN+SS) = 0.21 (note there is only one system of the class SN+SS H). Considering all multiple systems, the median difference of Hot - Warm is now $\Delta$[Fe/H] (SE Multi, SN Multi, SE+SN, SN+SS) = 0.17 $\pm$ 0.08 (MAD) dex.

\begin{figure}
\plotone{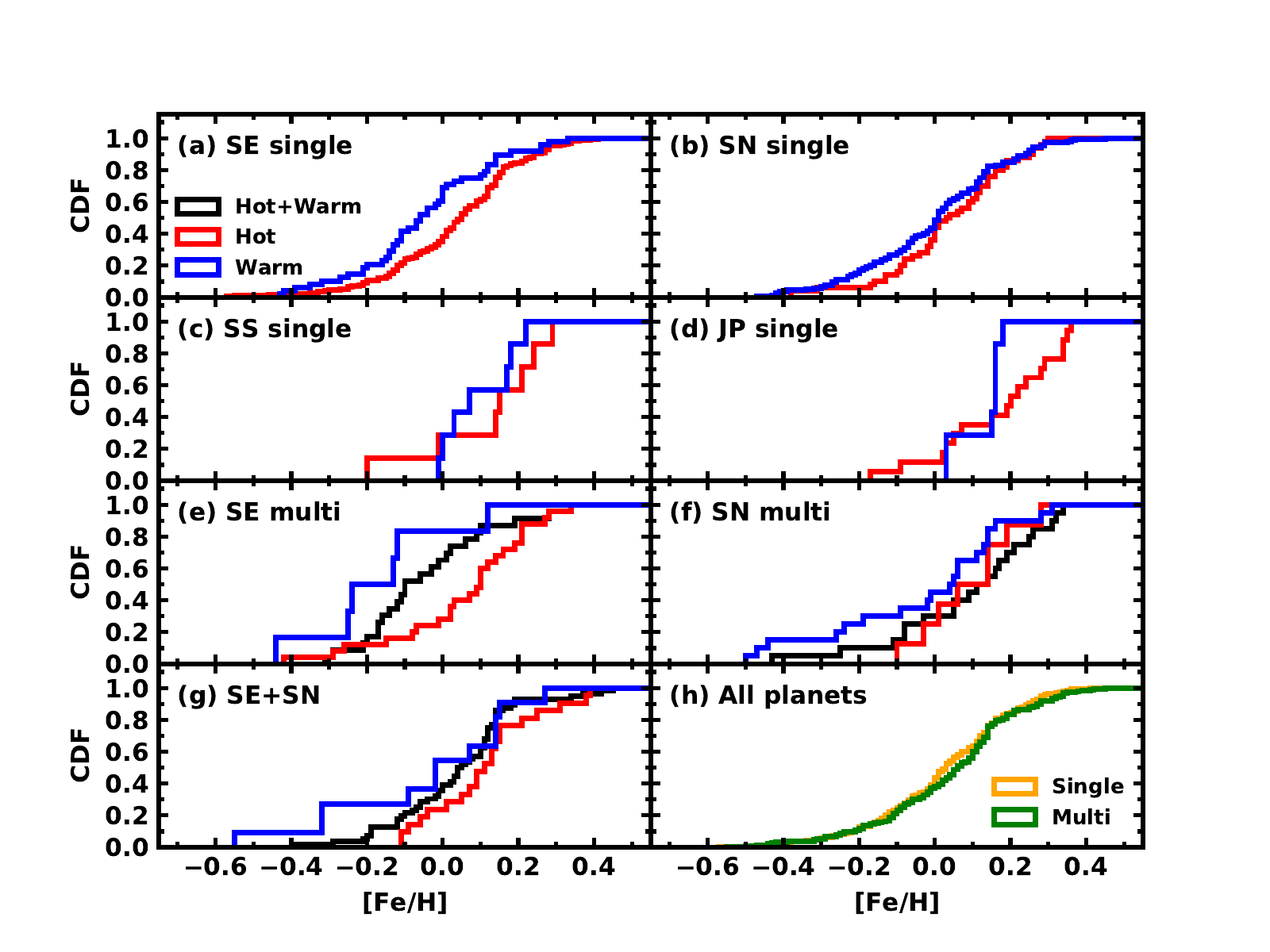}
\caption{Cumulative distribution functions (CDFs) for the metallicities of different samples: systems with a single SE ($R_{p} < $1.9 R$_{\oplus}$; panel a), systems with a single SN (1.9 R$_{\oplus} \leq R_{p} < $4.4 R$_{\oplus}$; panel b), systems with a single SS (4.4 R$_{\oplus} \leq R_{p} < $8.0 R$_{\oplus}$; panel c), systems with a single JP ($R_{p} \geq $8.0 R$_{\oplus}$; panel d), systems with multiple SE (panel e), systems with multiple SN (panel f), systems with SE+SN (panel g) and all systems (panel h). Systems with just hot, just warm or both hot and warm planets are represented by red, blue and black lines, respectively. In panel h, systems with single and multiple planets are shown by orange and green lines, respectively.}
\label{fig:feh_systems_cdfs}
\end{figure}

The hot-warm contrast in metallicity between the single and multiple planet systems can be further investigated using metallicity cumulative distribution functions (CDFs) for different planetary system subclasses. The different panels of Figure \ref{fig:feh_systems_cdfs} contain stellar host [Fe/H] CDFs for systems of SE Single, SN Single, SS single, JP single, SE multi, SN multi, and SE+SN, with each class separated into hot ($P \leq$ 10 d), warm (10 d $< P \leq$ 100 d), and hot+warm for those multiple systems containing both hot and warm planets. 
In panel (h) of the figure we also show the CDFs for our entire sample of all singles (orange CDF) and all multiples (green CDF). \cite{weiss2018} used CDFs to investigate possible differences in stellar properties between magnitude-limited CKS samples that host single versus multiple planets. They compared several stellar properties, including mass, [Fe/H], and projected rotational velocities, and found that the stellar properties were indistinguishable for the entire sample of single versus multiple systems. The same comparison using [Fe/H] CDFs was carried out here and we confirm the result from \cite{weiss2018} that the cumulative metallicity distribution of CKS stars hosting single planets is not measurably different from that of host stars of multiple planets: the Cucconi test gives $p_{CC} \approx 0.448$, indicating that we cannot reject the null hypothesis that singles and multiples, taken as whole, are drawn from the same parent population.


The general trend from the distributions in Figure \ref{fig:feh_systems_cdfs} is in line with the observation that hot planets are detected preferentially around more metal rich stars \citep[e.g.,][]{mulders2016,wilson2018,petigura2018} as we note that the red [Fe/H] CDFs (representing systems with hot planets) tend on average to be more metal rich when compared to the blue CDFs (systems with warm planets), but this metallicity signature may not be statistically significant in some cases. Focusing first on the host star metallicities of single systems, the shifts in the [Fe/H] CDF between hot and warm single Super-Earth systems is evident and this is in agreement with the 
results from Cucconi tests ($p_{CC} \approx 0.005$), which indicates that the hot and warm SE are probably from different parent populations, such that Super-Earths in close hot orbits are drawn from slightly more metal rich host stars. For single Sub-Neptunes, on the other hand, the differences between the hot and warm CDFs are not significant and the Cucconi tests between the [Fe/H] distributions give $p_{CC} \approx 0.259$, indicating that their parent populations are likely the same. 

For those systems with multiple Super-Earths, the CDFs for warm planets have a significant metal poor ([Fe/H] $<$ 0.00) portion; there is a large maximum metallicity displacement of roughly $\sim$0.3 dex for the CDF of hot versus warm SE Multi and this metallicity difference is more significant when compared to that seen for hot versus warm in the SE singles (Figure \ref{fig:feh_systems_cdfs}). Cucconi tests between the metallicity distributions of hot and warm SE multis give $p_{CC} \approx 0.046$, meaning that we cannot reject the null hypothesis that they are drawn from the same parent population. This is an intriguing result since it points to possible difference between systems with single and multiple Super-Earths. However, it should be kept in mind that there is a smaller number of systems with multiple planets.

For the Sub-Neptunes, the hot versus warm metallicity signature in the CDFs is not statistically significant. For the case of systems with multiple Sub-Neptunes the conclusions are similar to the SN single case: Cucconi tests between the metallicity distributions indicate that the parent distributions of hot and warm SN multis are probably indistinguishable ($p_{CC} \approx 0.456$).
For the case of systems having SE+SN planets, we also do not find any significant metallicity differences: $p_{CC} \approx 0.256$.
For systems hosting single Sub-Saturns or Jupiters, we also find that the [Fe/H] distributions are not significantly different: $p_{CC} \approx 0.420$ for SS and $p_{CC} \approx 0.043$ for JP.

In order to check our results, we performed the exact same analysis as above, but now using the sample, data ([Fe/H] and $R_{p}$) and radius boundaries for dividing the planets into different classes from \cite{petigura2018}. The conclusions are the same as the one from our tests: there is a significant difference between the metallicity distributions only when we compare systems with hot and warm single Super-Earths ($p_{CC} \approx 0.004$). For the single Sub-Neptunes, the results are also consistent with ours, but it is worth noting that the p-values are closer to 0.01 ($p_{CC} \approx 0.011$).

We checked whether the metallicity difference (H minus W) for Super-Earths might be the result of a mass-metallicity relation (i.e., more massive stars are overall more metal rich), although the mass-metallicity relation in the solar neighborhood is quite scattered. Stellar masses were investigated for the Super-Earths and it was found that the host stars of single SE hot planets are less massive by 0.03 $M_{\odot}$ (median) than those of that have single SE warm planets.
These host star mass comparisons indicate that, to first order, stellar mass by itself is not likely to be responsible for the differences. Host star masses in this CKS sample and their correlations with planet system properties will be presented and discussed in detail in a future paper.

While it has been shown by this and previous works that the stellar properties of stars hosting single and multiple systems are similar when considering the entire CKS sample, we find that there are some intriguing differences when the orbital periods are also taken into account. More specifically, we observe a statistically significant difference between the metallicity distributions for systems with hot and warm single Super-Earths, but a similar signature is not seen for stars hosting single sub-Neptunes and multiple Super-Earths and or Sub-Neptunes. We reiterate, however, that these findings are based upon planets as seen by Kepler, and as a result some of the systems may have misclassified architectures due to possible observational biases, such as, the low number of confirmed planets with periods larger than 100 days. If confirmed by further studies with larger and more complete samples of planetary systems, our results could point to an important difference within the regime of small planets which, to the best of our knowledge, currently lacks a theoretical explanation.
 
\section{Conclusions}
\label{sec:conclusions}

Paper I in this series \citep{martinez2019} presented a spectroscopic analysis of the California-{\it Kepler} Survey sample \citep{petigura2017a} and derived stellar radii with a median uncertainty of 2.8\%, and radii for 1633 planets, achieving an internal precision of 3.7\%, which provided constraints on the slope of the radius gap as a function of orbital period. The stellar radii determinations in Paper I required effective temperatures, which were derived --- along with surface gravities, microturbulent velocities ($\xi$), and stellar metallicities --- from a classical quantitative spectroscopic analysis using a sample of Fe I and Fe II lines. The set of Fe I and Fe II lines utilized derive from a strategic selection of spectral lines spanning a range of excitation potentials and typical line strengths that provide accurate values of $T_{\rm eff}$, $\log{g}$, $\xi$, and [Fe/H] \citep{ghezzi2010a,ghezzi2018}. The independent spectroscopic analysis employed here on the CKS sample results in values of [Fe/H] with typical internal uncertainties of $\pm$0.02 dex, that compare well with other results from the literature, including those from the SDSS-IV APOGEE and LAMOST surveys, with median differences within less than 0.10 dex. 

Using a ``clean'' sample of 961 planets orbiting 663 stellar hosts, we make a number of comparisons between stellar host metallicities and their planet characteristics, as well as planetary system architectures. For the various comparisons, planets and planetary systems were divided into broad groups: four classes based on planetary radius (Super-Earths, Sub-Neptunes, Sub-Saturns, Jupiters), two orbital period bins (``hot'', with $P$ $\leq$ 10 d, and ``warm'', with 10 d $< P \leq$ 100 d), and whether the system has a single or multiple transiting planets (``single'' and ``multi''), while it is understood that both single and multiple systems may contain additional planets not detected by \textit{Kepler}. All discussions about host star metallicities, planets, and planetary systems are restricted to the studied sample of CKS stars and planets with periods of less than 100 days. The main results from this study are summarized below.

\bigskip
1) The metallicity distribution of the host stars of Super-Earths (SE, R$_{p}$ $<$1.9 R$_{\oplus}$) is found to be very similar to the metallicity distribution of Galactic disk red giant stars in the solar neighborhood with Galactocentric distances between 7-9 kpc \citep{hayden2015} as explored by the APOGEE survey \citep{majewski2017}. The progenitors of such APOGEE red giants are typically FGK main-sequence stars that overlap with the population studied in the CKS survey and analyzed here. 
The close kinship of SE hosts with the Galactic solar neighborhood distribution may suggest that the underlying processes governing the formation of SE are not, in general, strong functions of host star metallicity; this does not imply, however, that metallicity does not affect certain aspects of small-planet formation, such as short-period versus long-period systems, or single versus multiple planet systems as we highlight below. 

\bigskip
2) For the case of the Sub-Neptunes (SN) only the smallest SN in the studied sample have metallicities that are similar to the fiducial Galactic metallicity distribution. 
Over the range of the radii spanned by the SN class, the Cucconi p-values are rapidly decreasing as the distributions become increasingly divergent at roughly R$_{p}\sim$ 2.5 R$_{\oplus}$.
This divergence is possibly driven by the strong metallicity dependence of the Sub-Neptunes occurrence rates \citep{petigura2018,owen2018}.   
At ever-increasing planetary radii, the Cucconi p-values reach a minimum value and then become nearly constant, with only a small decrease out to R$_{p}\sim$8 R$_{\oplus}$.  The transition from the rapidly-evolving metallicity distribution function, to a much slower change with planetary radius occurs over the interval of R$_{p}\sim$ 3.9 -- 4.9 R$_{\oplus}$ and we interpret this radius interval as the boundary between Sub-Neptunes (SN) and Sub-Saturns (SS) with the middle of the interval occurring at 4.4 $\pm$ 0.5 R$_{\oplus}$.

\bigskip
3) The derived [Fe/H] distributions confirm the well-known observation that large planets (SS and JP) are found around host stars whose metallicity distribution is significantly skewed towards super-solar metallicities \citep[e.g.,][]{sousa2008,ghezzi2010a,petigura2018}. The median host-star metallicity of the SS and JP is [Fe/H] = +0.17. For the sample analyzed in this study, there seems to be a low-metallicity cut-off at about [Fe/H]$\sim$-0.2, below which no Sub-Saturn and Jupiter host stars are found. 

\bigskip
4) The behavior of host-star median-metallicities between hot and warm subclasses in single-planet and multiple-planet systems, respectively, were found to be different. Within the same subclasses in single systems, the differences in median metallicity values between hot and warm systems are found to be relatively small, at a level of 0.00 to 0.10 dex across SE single, SN single, SS single, and JP single subclasses, with a mean difference of 0.06 dex. Multiple-planet systems present significant differences in median host-star Fe abundances between hot and warm systems, with stars hosting hot multiple-planet systems being, on average, more metal-rich than the warm multiple-planet system host stars by 0.17 dex, averaged over all planet size classes.

\bigskip
5) Stellar host [Fe/H] cumulative distribution functions (CDFs) were investigated for single and multiple small planet systems (SE, SN, and SE+SN), as well as singles versus multiples in general. A comparison of the metallicity CDFs between systems having single versus multiple planets, with no further subdivision, finds no significant differences between their parent-star metallicity distributions, in agreement with previous results in the literature \citep{weiss2018,munozromero2018}. However, when the single and multiple planet systems containing only Super-Earths are divided into hot and warm groups, we observe small, but significant, metallicity CDF shifts between the hot and warm populations; host-star metallicities of single systems with hot SE planets are drawn from slightly more metal-rich host stars. In systems having multiple SE, the differences in the metallicity CDFs for hot versus warm are larger still.
In contrast, we are unable to measure a statistically significant metallicity signature for the hot and warm populations of systems with both single and multiple SN. 

\acknowledgments
{We thank the referee for suggestions that improved the paper significantly. We also thank the AAS Journal statistics editor for suggesting the use of more powerful statistical tests. We warmly thank the California Kepler Survey team for making their Keck HIRES reduced data available. We thank Michael Hayden for sending the original data used for defining the APOGEE solar neighborhood MDF and Fabio Wanderlei for doing comparisons between APOGEE DR12 and DR16. We thank Hilke Schlichting and Yanqin Wu for insightful discussions. L.G. would like to thank the financial support from Fundação de Amparo à Pesquisa do Estado do Rio de Janeiro (FAPERJ) through the ARC grant number E-26/211.386/2019. The Guoshoujing Telescope (used for the Large Sky Area Multi-Object Fiber Spectroscopic Telescope LAMOST) is a National Major Scientific Project built by the Chinese Academy of Sciences. Funding for the project has been provided by the National Development and Reform Commission. LAMOST is operated and managed by the National Astronomical Observatories, of the Chinese Academy of Sciences. This work has made use of the VALD database, operated at Uppsala University, the Institute of Astronomy RAS in Moscow, and the University of Vienna.}



\software{ARES \citep{sousa2015}, 
         {Astropy \citep{astropy:2013, astropy:2018}}, 
         {IRAF \citep{1986SPIE..627..733T,1993ASPC...52..173T}}, 
         {MOOG \citep{sneden1973}}, 
         {matplotlib \citep{Hunter:2007}}, 
         {numpy \citep{2020NumPy-Array}}, 
         {pandas \citep{mckinney-proc-scipy-2010}}, 
         {scipy \citep{2020SciPy-NMeth}}
         }





\bibliographystyle{yahapj}
\bibliography{references}




\end{document}